\pdfoutput=1

\documentclass[11pt,twoside,a4paper,cmspaper,final,collab]{cms-tdr}
\def\svnVersion{317839}\def\cmsCernNo{2013-037}\def\cmsCernDate{\today}\def\cmsMessage{Published in the European Physical Journal C as \href{http://dx.doi.org/10.1140/epjc/s10052-015-3853-3}{\doi{10.1140/epjc/s10052-015-3853-3.}}}

\begin{document}\cmsNoteHeader{HIG-14-016}

\hyphenation{had-ron-i-za-tion}
\hyphenation{cal-or-i-me-ter}
\hyphenation{de-vices}
\RCS$Revision: 312958 $
\RCS$HeadURL: svn+ssh://svn.cern.ch/reps/tdr2/papers/HIG-14-016/trunk/HIG-14-016.tex $
\RCS$Id: HIG-14-016.tex 312958 2015-12-02 02:21:28Z alverson $
\newlength\cmsFigWidth
\ifthenelse{\boolean{cms@external}}{\setlength\cmsFigWidth{0.85\columnwidth}}{\setlength\cmsFigWidth{0.4\textwidth}}
\ifthenelse{\boolean{cms@external}}{\providecommand{\cmsLeft}{top}\xspace}{\providecommand{\cmsLeft}{left}\xspace}
\ifthenelse{\boolean{cms@external}}{\providecommand{\cmsRight}{bottom}\xspace}{\providecommand{\cmsRight}{right}\xspace}

\newcommand{\MADGRAPHAMCATNLO} {{\textsc{madgraph5}\_{a}\textsc{mc@nlo}}\xspace}
\newcommand{\POWHEGMINLO} {{\textsc{powheg+minlo}}\xspace}
\newcommand{\HRES} {{\textsc{hres}}\xspace}

\newcommand{\statdot}{\ensuremath{\,\text{(stat.)}}\xspace}
\newcommand{\systdot}{\ensuremath{\,\text{(syst.)}}\xspace}

\newcommand{\mgg}{\ensuremath{m_{\gamma\gamma}}\xspace}
\newcommand{\mjj}{\ensuremath{m_\mathrm{jj}}\xspace}
\newcommand{\mH}{\ensuremath{m_{\PH}}\xspace}
\newcommand{\ptg}{\ensuremath{\pt^{\gamma}}\xspace}
\newcommand{\ptj}{\ensuremath{\pt^\mathrm{j}}\xspace}
\newcommand{\ptga}{\ensuremath{\pt^{\gamma1}}\xspace}
\newcommand{\ptgb}{\ensuremath{\pt^{\gamma2}}\xspace}
\newcommand{\ptja}{\ensuremath{\pt^\mathrm{j1}}\xspace}
\newcommand{\ptgg}{\ensuremath{p_{\mathrm{T}}^{\gamma\gamma}}\xspace}
\newcommand{\Hgg}{\ensuremath{\PH\to\Pgg\Pgg}\xspace}
\newcommand{\gee}{\ensuremath{\gamma\to\Pep\Pem}\xspace}
\newcommand{\Zee}{\ensuremath{\cPZ\to\Pep\Pem}\xspace}
\newcommand{\Zmm}{\ensuremath{\cPZ\to\Pgmp\Pgmm}\xspace}
\newcommand{\Zmmg}{\ensuremath{\cPZ\to\Pgmp\Pgmm\Pgg}\xspace}
\newcommand{\Wenu}{\ensuremath{\PW\to\Pe\Pgn}\xspace}
\newcommand{\ttH}{\ensuremath{\cPqt\cPaqt\PH}\xspace}
\newcommand{\HqT}{{\textsc{h}q\textsc{t}}\xspace}
\newcommand{\costhetastar}{\ensuremath{\cos\theta^{\ast}}\xspace}
\newcommand{\twomp}{\ensuremath{2^{+}_{m}}\xspace}
\newcommand{\Njets}{\ensuremath{N_\text{jets}}\xspace}
\newcommand{\ygg}{\ensuremath{\abs{y^{\gamma\gamma}}}\xspace}
\newcommand{\yggj}{\ensuremath{\abs{y^{\gamma\gamma}-y^{\mathrm{j1}}}}\xspace}
\newcommand{\deltaphigg}{\ensuremath{\Delta \phi^{\gamma\gamma}}\xspace}
\newcommand{\deltaphijj}{\ensuremath{\Delta \phi^{\mathrm{jj}}}\xspace}
\newcommand{\deltaetajj}{\ensuremath{\Delta \eta^{\mathrm{jj}}}\xspace}
\newcommand{\deltaphiggjj}{\ensuremath{\Delta \phi^{\gamma\gamma,\mathrm{jj}}}\xspace}
\newcommand{\Zepp}{\ensuremath{\abs{\eta^{\gamma\gamma}-(\eta^{\mathrm{j1}}+\eta^{\mathrm{j2}})/2}}\xspace}

\cmsNoteHeader{HIG-14-016}
\title{Measurement of differential cross sections for Higgs boson production in the diphoton decay channel in pp collisions at $\sqrt{s}=8\TeV$}
\titlerunning{Higgs production differential cross sections at $\sqrt{s}=8$\TeV}

\date{\today}

\abstract{
A measurement is presented of differential cross sections for Higgs boson (H) production in pp collisions at $\sqrt{s}=8$\TeV.
The analysis exploits the $\PH\to\Pgg\Pgg$ decay in data corresponding to an integrated luminosity of 19.7\fbinv collected by the CMS experiment at the LHC.
 The cross section is measured as a function of the kinematic properties of the diphoton system and of the associated jets.
 Results corrected for detector effects are compared with predictions at next-to-leading order and next-to-next-to-leading order in perturbative quantum chromodynamics, as well as with predictions beyond the standard model.
For isolated photons with pseudorapidities $\abs{\eta}<2.5$, and with the photon of largest and next-to-largest transverse momentum ($\pt^{\gamma}$) divided by the diphoton mass $m_{\gamma\gamma}$ satisfying the respective conditions of $\pt^{\gamma}/m_{\gamma\gamma}> 1/3$ and ${>}1/4$, the total fiducial cross section is $32 \pm 10$\unit{fb}.}

\hypersetup{%
pdfauthor={CMS Collaboration},%
pdftitle={Measurement of differential cross sections for Higgs boson production in the diphoton decay channel in pp collisions at sqrt(s)=8 TeV},%
pdfsubject={CMS},%
pdfkeywords={Higgs, differential cross section, diphoton, QCD}}

\maketitle

\section{Introduction}
\label{sec:Introduction}

In 2012, the ATLAS and CMS Collaborations announced the observation~\cite{Aad:2012tfa, Chatrchyan:2012ufa, Chatrchyan:2013lba} of a
new boson with a mass of about 125\GeV, with properties consistent with expectations for the
standard model (SM) Higgs boson.
The Higgs boson (\PH) is the particle predicted to exist as a
consequence of the spontaneous symmetry breaking mechanism acting in the electroweak sector of the
SM~\cite{Glashow:1961tr,Weinberg:1967tq,sm_salam}.
This mechanism was suggested more than fifty years ago,
and introduces a complex scalar field, which gives masses to W and Z bosons ~\cite{Englert:1964et,Higgs:1964pj,Guralnik:1964eu,Higgs:1966ev,Kibble:1967sv}.
The scalar field also gives mass to the
fundamental fermions through a Yukawa interaction \cite{Weinberg:1967tq}.
Couplings and spin of the new boson are found to be consistent with the SM predictions \cite{HggLegacy,Aad:2014eha,CMSHiggsCombLegacy,Aad:2013xqa,Khachatryan:2014kca}.
A measurement of the $\Pp\Pp \to \Hgg$ differential cross section as a function of kinematic observables investigates possible deviations in distributions related to production, decay, and additional jet activity. It provides a check of perturbative calculations in quantum chromodynamics (QCD), and can point to alternative models in the Higgs sector. A similar analysis has been carried out by the ATLAS Collaboration in diphoton and four-lepton decay channels \cite{Aad:2014lwa, Aad:2014tca, Aad:2015lha}.

Despite its small
branching fraction of $\approx$0.2\%~\cite{LHCHiggsCrossSectionWorkingGroup:2013} predicted by the SM, the
\Hgg decay channel provides a clean final-state topology and a precise reconstruction of the diphoton mass.
The dominant background arises from irreducible direct-diphoton production and
from the reducible $\Pp\Pp \to \gamma$+jets and
$\Pp\Pp \to \text{jets}$ final states.
The relatively high efficiency of the \Hgg selection makes this final state one of the most important channels for
observing and investigating the properties of the new boson.

In this paper, the cross section is measured as a function of the kinematic properties of the diphoton system, and, in events with at least one or two accompanying jets, also as a function of jet-related observables.
Two isolated photons are required to be within pseudorapidities $\abs{\eta}<2.5$, and the photon with largest and next-to-largest transverse momentum ($\ptg$) must satisfy the respective conditions of $\ptg/\mgg>1/3$ and ${>}1/4$, where \mgg represents the diphoton mass.
The transverse momentum $\ptgg$ and the rapidity $\ygg$ of the Higgs boson, observables related to the opening angle between the two photons, and the number of jets $\Njets$ with $\pt > 25$\GeV produced in association with the diphoton system, are defined in this inclusive fiducial selection. A departure relative to the SM-predicted angular distributions would be an important observation, as it could reflect different spin and parity properties \cite{Gao:2010qx} than expected in the SM.

The variables defined with at least one accompanying jet are sensitive to the transverse Lorentz boost of the diphoton system. A modification in the corresponding distributions or in the $\ptgg$ spectrum could signify new contributions to gluon-gluon fusion production of the Higgs boson (ggH) \cite{GrojeanHiggsPt}.
The variables defined by requiring at least two accompanying jets are related to production of \Hgg through vector boson fusion (VBF); however, given the low event yield after selecting two jets, no other selection is applied to enhance this production mechanism.
This is different from what was done in Ref.~\cite{HggLegacy}, where an attempt was made to classify the events according to the production mechanism. The differential cross sections are therefore mainly sensitive to the dominant ggH production mode of the Higgs boson.

The data correspond to an integrated luminosity of 19.7\fbinv collected at the CERN LHC by the CMS experiment in proton-proton collisions at $\sqrt{s}=8\TeV$.
The trigger requirements and vertex determination are identical to those of Ref.~\cite{HggLegacy}, while photon selection and event classification are modified to reduce their dependence on $\ptg$ and $\eta^{\gamma}$, providing thereby a less model-dependent measurement.
Photons are identified using a multivariate classifier that combines information on distributions of shower and isolation variables designed to be independent of $\ptg$ and $\eta^{\gamma}$.
The signal yield is extracted by fitting the $\mgg$ distribution simultaneously in all bins of the observables.
To improve the sensitivity of the analysis, the selected events are categorized using an estimator of the mass resolution that is not correlated with \mgg, which simplifies the description of the background.
Measured distributions are unfolded for detector effects and compared to distributions at the generator level from the latest Monte Carlo (MC) predictions.

The paper is organized as follows. After a brief description of the CMS detector and event reconstruction given in Section~\ref{sec:CMS}, and of the simulated samples in Section~\ref{sec:samples}, the photon selection and event classification are detailed in Section~\ref{sec:selection}, where we also describe the kinematic observables.
Section~\ref{sec:unfolding} provides the statistical methodology for extracting the signal, and gives details on modelling signal and background, and on the unfolding procedure.
Systematic uncertainties are detailed in Section~\ref{sec:systematics}.
Unfolded results are then compared with theoretical predictions in Section~\ref{sec:theory}, and a brief summary is given in Section~\ref{sec:conclusions}.

\section{The CMS detector}
\label{sec:CMS}
A full description of the CMS detector, together with a definition of the coordinate system and the relevant kinematic variables, can be found in Ref.~\cite{Chatrchyan:2008zzk}.
Its central feature is a superconducting solenoid,
13\unit{m} in length and 6\unit{m} in diameter,
which provides an axial magnetic field of 3.8\unit{T}.
The core of the solenoid is instrumented with trackers and calorimeters.
The steel flux-return yoke outside the solenoid is equipped with gas-ionisation detectors
used to reconstruct and identify muons.
Charged-particle trajectories are measured using silicon pixel and
strip trackers, within $\abs{\eta} < 2.5$.
A lead tungstate crystal electromagnetic calorimeter (ECAL) and a
brass and scintillator
hadron calorimeter (HCAL) surround the tracking volume and cover the region
$\abs{\eta} < 3$.
The ECAL barrel extends to $\abs{\eta} < 1.48$, while the ECAL endcaps cover the region $1.48 < \abs{\eta} < 3.0$.
A lead and silicon-strip preshower detector is located in front of each
ECAL endcap in the region $1.65 < \abs{\eta} < 2.6$.
The preshower detector includes two planes of silicon sensors that measure the
transverse coordinates of the impinging particles.
A steel and quartz-fibre Cherenkov calorimeter extends the coverage to $\abs{\eta} < 5.0$.
In the $(\eta, \phi)$ plane, for $\abs{\eta} < 1.48$, the HCAL cells map onto $5\times5$
ECAL crystal arrays to form calorimeter towers projecting radially
outwards from points slightly offset from the nominal interaction point.
In the endcap, the ECAL arrays matching the HCAL cells contain fewer crystals.
To optimize the energy resolution, the calorimeter signals are
calibrated and corrected for several detector effects~\cite{Chatrchyan:2013dga}.
Calibration of the ECAL uses the $\phi$-symmetry of the energy flow, photons from $\pi^{0}\to\gamma\gamma$ and $\eta\to\gamma\gamma$, and electrons from \Wenu and \Zee decays.
Changes in the transparency of the ECAL crystals due to irradiation
during the LHC running periods and their subsequent recovery
are monitored continuously
and corrected, using light injected from a laser system.

Photons are reconstructed from clusters of energy deposition
into so-called ``superclusters'' in the ECAL \cite{Khachatryan:2015iwa}.
Events are selected using triggers requiring two photons
with different thresholds in energy in the transverse plane ($\ET$), respectively, $\ET>26$ and $>$18\GeV for the leading and subleading photons, and through other complementary selections.
One selection requires a loose calorimetric identification based on the distribution in energy
in the electromagnetic cluster, and loose isolation requirements on photon candidates.
The other selection requires a photon candidate to have a high value of
\RNINE~variable, defined as the sum of the energies deposited in the array of
$3{\times}3$ crystals centred on the crystal with highest energy deposition in the
supercluster, divided by the energy of the supercluster.
Photons that convert to $\Pep\Pem$ pairs before reaching the calorimeter tend to have wider showers and smaller values of \RNINE\
than unconverted photons.
High trigger efficiency is maintained by having both photons satisfy either selection.
The measured trigger efficiency is 99.4\% for events satisfying the diphoton preselections described in Section~\ref{sec:selection}.

\section{Monte Carlo samples}
\label{sec:samples}

The MC simulation of detector response employs a detailed description of the CMS detector,
and uses \GEANTfour version~9.4.p03~\cite{Agostinelli:2002hh}.
Simulated events include additional $\Pp\Pp$ collisions that take place in or close to the time span of the bunch crossing, and overlap the interaction of interest.
The probability distribution of these pileup events is weighted to reproduce the observed number of interactions in data.

{\tolerance=600
The MC samples for ggH and VBF processes use the next-to-leading order (NLO)
matrix element generator \POWHEG (version 1.0)~\cite{powheg1,powheg2,powheg3,powheg-ggH,powheg-VBF}
interfaced with \PYTHIA 6.426 \cite{Sjostrand:2006za}.
The CT10 \cite{CT10nlo} set of parton distribution functions (PDF) is used in the calculation.
The \POWHEG generator is tuned following the recommendations of Ref.~\cite{LHCHiggsCrossSectionWorkingGroup2} and reproduces the Higgs boson $\pt$ spectrum predicted by the \HqT calculation \cite{HqT1, HqT2}.
The \PYTHIA6 tune Z2* is used to simulate the hadronization and underlying event in pp collisions at 8\TeV.
The Z2* tune is derived from the Z1 tune \cite{Field:2010bc}, which uses the CTEQ5L PDF, whereas Z2* adopts CTEQ6L1 \cite{Pumplin:2002vw}.
The cross section for the ggH process is reduced by 2.5\% for all values of \mH to
accommodate its interference with nonresonant diphoton production~\cite{Dixon:2003yb}.
The \PYTHIA6 generator is used alone for the VH (where V represents either the W or Z boson) and \ttH processes with the CTEQ6L1 PDF \cite{Pumplin:2002vw} and Z2* tune.
The SM cross sections and branching fractions are taken
from Ref.~\cite{LHCHiggsCrossSectionWorkingGroup:2013}.
\par}

The samples of Drell--Yan events ($\qqbar\to \Z/\gamma^{*} \to \ell^{+}\ell^{-}$, where $\ell$ is a lepton), and background samples used to represent the diphoton continuum and processes where one of the photon candidates arises from misidentified jet fragments, are the same as used in Ref.~\cite{HggLegacy}.
Simulated samples of \Zee, \Zmm, and \Zmmg events, used for comparison with data and to extract an energy scale and corrections for resolution of photon energies, are generated with \MADGRAPH, \SHERPA, and \POWHEG~\cite{powheg-Zjj}, providing comparisons among the different generators. Simulated background samples are used for training of multivariate discriminants, and for defining selection and classification criteria.

{\tolerance=1800
The diphoton continuum processes involving two prompt photons are simulated
using \SHERPA 1.4.2~\cite{Gleisberg:2008ta}.
 The remaining processes where one of the photon candidates arises from misidentified jet fragments are simulated using \PYTHIA6 alone.
\par}

A comparison of unfolded data with results from models for ggH using the MC generators \HRES \cite{HRes1,HRes2}, \POWHEG and \POWHEGMINLO \cite{PowhegMinlo}, and \MADGRAPHAMCATNLO \cite{aMC@NLO} is presented in Section~\ref{sec:theory}.
\section{Event selection and classification}
\label{sec:selection}

{\tolerance=1200
Trigger requirements, vertex determination, and kinematic criteria on photons are unchanged relative to those given in Ref.~\cite{HggLegacy}.
The multivariate classifiers used to identify photons and to estimate mass resolution are also unchanged, but used in a different way.
Instead of using a discriminant for photon identification as an input to the final diphoton kinematic discriminant, a requirement is set on the photon identification discriminant.
Event classification, instead of being based on the output of the kinematic discriminant, is based on the estimated $\mgg$ resolution. These differences are described in greater detail in the following section.
\par}

\subsection{Photon identification}
\label{sec:photonId}

Photon candidates are required to be within the fiducial region of $\abs{\eta}<2.5$, excluding the barrel--endcap transition region of $1.44 < \abs{\eta} < 1.57$, where photon reconstruction is not optimal.
The transverse momenta of the two photons are required to satisfy the previously mentioned conditions of $\ptga/\mgg>1/3$ and $\ptgb/\mgg>1/4$. The use of $\pt$ thresholds scaled by $\mgg$ prevents the distortion of the low end of the $\mgg$ spectrum that results if a fixed threshold is used.
Photons are also required to satisfy preselection criteria based on isolation and distributions in shower variables slightly more stringent than used in the trigger requirements.

Three variables are calculated for each reconstructed candidate for a $\Pp\Pp$ interaction vertex:
the sum of the $\pt^2$ of the charged-particle tracks emerging from the vertex,
and two variables that quantify the difference in the vector and scalar sums in \pt between the diphoton system and the charged-particle tracks associated with the vertex.
In addition, if either photon is associated with any charged-particle track identified as resulting from $\gee$ conversion, extrapolation of their trajectories is used to clarify the origin of the vertex of their production.
These variables are used as inputs to a multivariate system based on a boosted decision tree (BDT) classifier to choose the reconstructed vertex to associate with the diphoton system.
All BDTs are implemented using the \textsc{tmva}~\cite{Hocker:2007ht} framework.

{\tolerance=800
Another BDT is trained to separate prompt photons from photon candidates resulting from misidentification of jet fragments passing the preselection requirements.
Inputs to the BDT are variables related to the lateral spread of the shower, and isolation energies reconstructed from scalar sums in $\pt$ of charged particles and $\ET$ sums of photons in a cone with an opening angle $\Delta R = \sqrt{\smash[b]{(\Delta \eta )^{2}+(\Delta\phi)^{2}}} <0.3$ around the photon, computed using the particle-flow (PF) algorithm \cite{CMS-PAS-PFT-09-001, CMS-PAS-PFT-10-001}.
\par}

\begin{figure}[htb]
 \centering
   \includegraphics[width=0.45\textwidth]{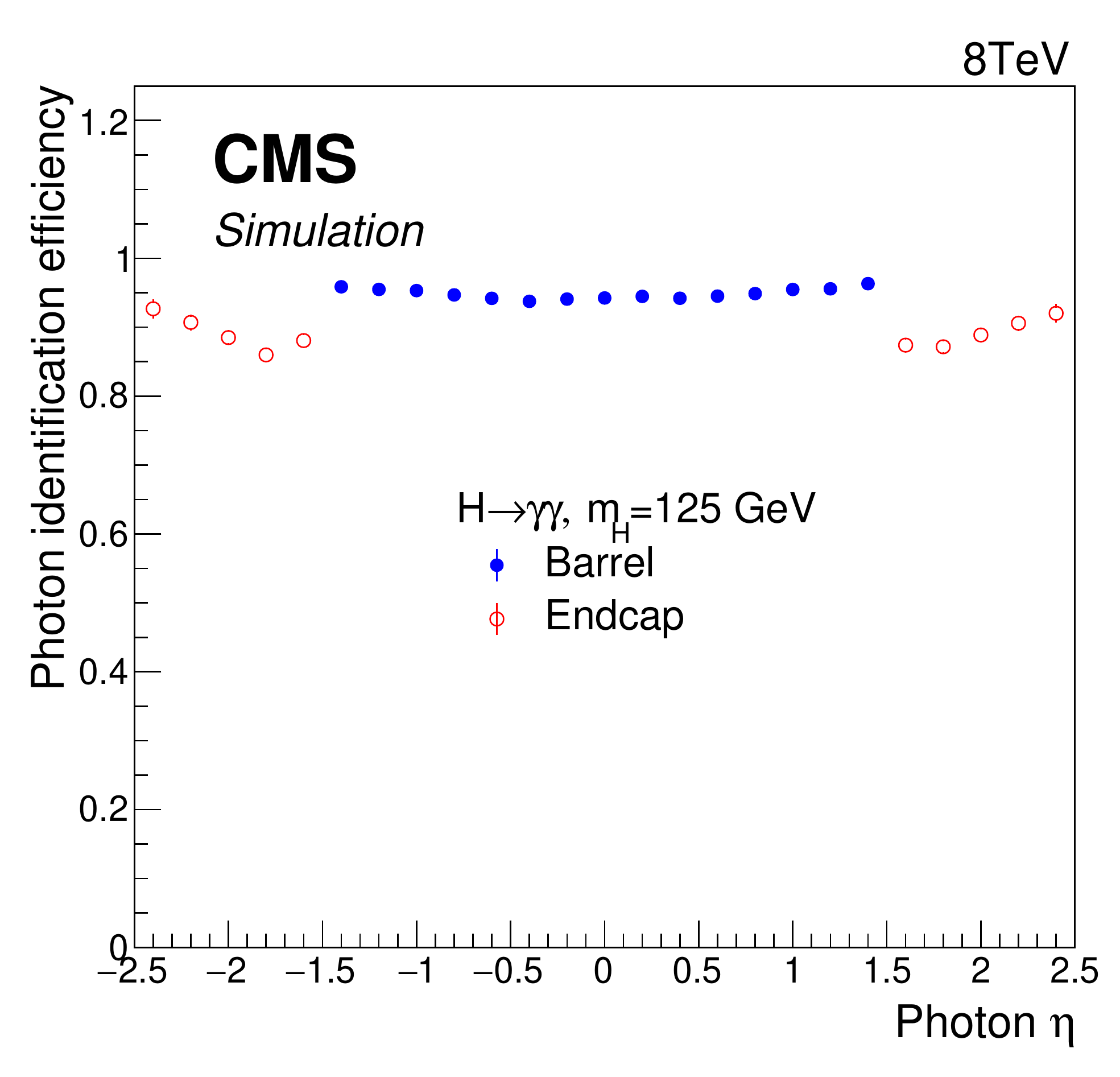}
   \includegraphics[width=0.45\textwidth]{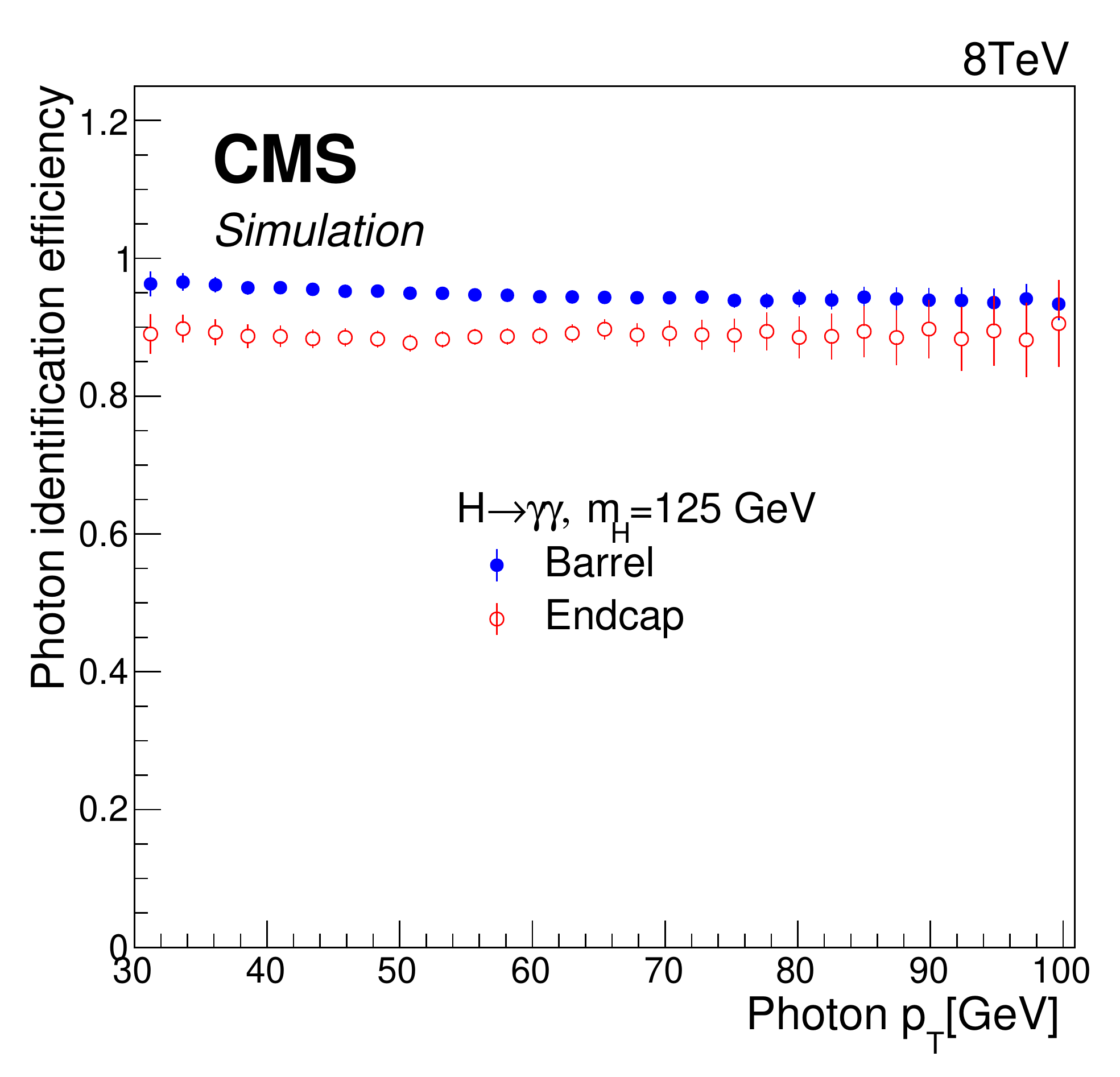}
   \caption{Photon identification efficiencies for a Higgs boson with $\mH=$125\GeV, as a function of photon pseudorapidity (\cmsLeft), and photon transverse momentum (\cmsRight).}
  \label{fig:PhoIdEff}

\end{figure}

The $\eta$ and energy of the supercluster corresponding to the reconstructed photon are also included as input variables in the photon identification BDT. These variables are introduced to explicitly correlate the shower topology and isolation variables with $\eta$ and \pt.
Furthermore, during the BDT training, the $\eta$ and \pt background distributions are reweighted to match the distributions in the signal.
As a result, for a given requirement on the BDT output, efficiencies for photon identification are almost independent of $\eta$ and \pt, as can be seen in Fig.~\ref{fig:PhoIdEff},
 which provides less model-dependent efficiency corrections for comparison of data and MC expectations at the particle level.
A loose selection is applied on the BDT output for photons detected in the barrel region, while a tight selection is applied to endcap photons, with respective mean efficiencies of 95\% and 90\% for signal photons.

Photon efficiencies in signal samples are corrected for the difference in efficiency between data and simulation as measured with \Zee events, treating electrons as photons by reweighting the electron cluster variable \RNINE~to that of the \RNINE~distribution of signal photons.
 There is good agreement found in the BDT output between data and simulation in \Zee and \Zmmg events.

\subsection{Event classification using an estimator of mass resolution}
\label{sec:evClasses}

Measuring differential kinematic distributions implies that events cannot be classified according to the diphoton kinematic BDT used for the inclusive measurement of Ref.~\cite{HggLegacy}, as that would create a bias in the result.
To improve the performance beyond the simple classification using the \RNINE~variable in the reference sequential analysis, referred to as ``cut-based analysis" in Ref.~\cite{HggLegacy}, an event categorization is introduced that is based on the estimated energy resolution.

The photon energies are corrected using a multivariate regression technique for the containment of showers in clustered crystals, for shower losses of photons that convert in the material upstream of the calorimeter, and for effects from pileup, based on shower variables and variables related to positions of photons in the detector studied in $\gamma$+jets simulated events.
The energy response to photons is parameterized through an extended form of the Crystal Ball function~\cite{CrystalBall} with a Gaussian core and two power law contributions.
The regression provides an estimate of the parameters of the function, and therefore a
prediction for the distribution in the ratio of the uncorrected supercluster energy to the true energy.
The correction to the photon energy is taken as the inverse of the most probable value of this distribution.
The standard deviation of the Gaussian core provides an estimate of the uncertainty in the energy ($\sigma_E$).

We define the estimator of the diphoton mass resolution by:
\begin{equation}
 \frac{\sigma_m}{\mgg} = \frac{1}{2} \sqrt{ \Big( \frac{\sigma_{E_1}}{E_1} \Big)^2 + S_1^2 + \Big(\frac{\sigma_{E_2}}{E_2} \Big)^2 + S_2^2}
\end{equation}
where $E_1$ and $E_2$ are the corrected energies of the two photons from the regression, $\sigma_{E_1}$ and $\sigma_{E_2}$ are the uncertainties in the photon energies,
and $S_1$ and $S_2$ are smearing terms depending on $\eta$, $\RNINE$, and $\ET$, determined from \Zee events, needed for the simulation to match the energy resolution in data.

For the typical energy range of photons in this analysis, the relative energy resolution $\sigma_E/E$ depends on the energy, which in turn introduces a dependence of the mass resolution on the value of \mgg.
Therefore, a categorization based on the diphoton mass resolution introduces a distortion of the background mass spectrum.
To obtain a smooth background description, we apply a transformation to the estimator $\sigma_E$ to decorrelate $\sigma_E/E$ from the energy.

The value of $\sigma_E/E$ depends on $\eta$ because of differences in material in front of the ECAL and the inherent properties of the ECAL.
To ensure that the decorrelation is performed independent of the $\eta$ distribution of the training sample, in a first step a transformation is applied to make $\sigma_E/E$ independent of $E$ and $\eta$.
A $\gamma$+jets MC sample is used to build the fully decorrelated variable, that covers a wide range in $\eta$ and $\pt$.
The decorrelation is performed by making a change of variable, replacing the probability distribution in $\sigma_E/E$ by its cumulative distribution function $cdf(\sigma_E/E)$.
This function follows a uniform distribution \cite{cowan1998statistical} and removes any correlation between $\sigma_E/E$ and ($E$,$\eta$).
Since the $\eta$ dependence is removed, this variable is not suitable for estimating a per-event mass resolution, which is non-uniform in the detector.
A second step is therefore introduced to restore the correlation of the energy resolution with photon $\eta$ in a \Hgg MC sample with $\mH=123$\GeV (statistically independent of the simulated events used to model the signal), and recover thereby a dependence of $\sigma_E$ on $\eta$ for the photons of interest.
In this step a new change of variable is performed, replacing the previous $\sigma_E/E$ with the inverse of the cumulative distribution function of $\sigma_E/E(\eta)$.
The impact of the particular $\pt$ spectrum used for this step is only modest for the correlation of $\sigma_E$ with $\eta$, which is in any case dominated by the material distribution in front of the ECAL and by calorimeter performance.
The advantage of such a two-step procedure is that it offers a decorrelated $\sigma_E/E$ variable that is uncorrelated with $E$ and does not depend on the $\eta$ distribution of the training sample.
It provides the typical energy resolution at a given $\eta$ of the detector.
The final result can be interpreted as an estimator of the average energy resolution at a particular value of $\eta$.

\begin{figure}[!htb]
 \centering
   \includegraphics[width=0.45\textwidth]{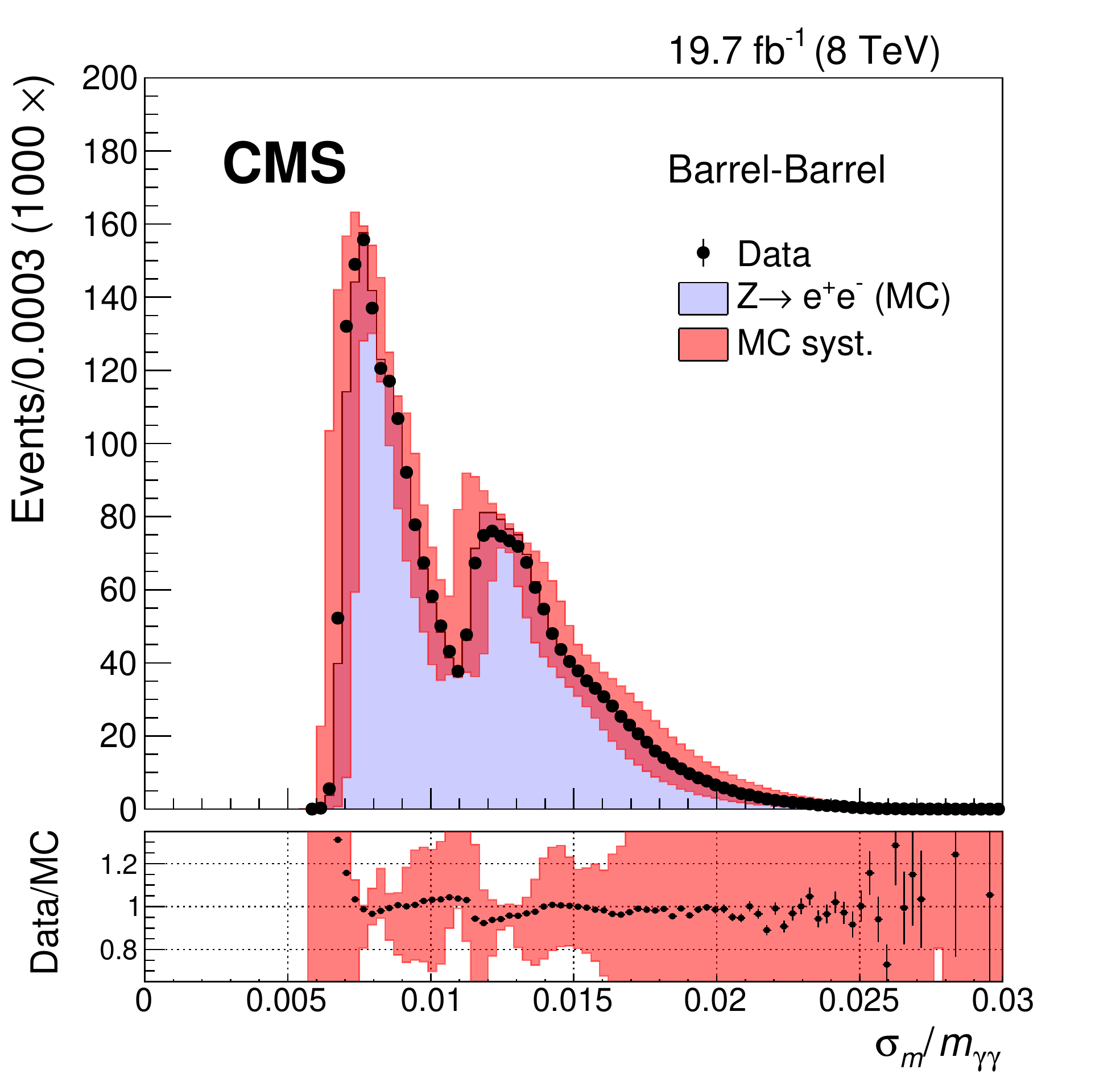}
   \includegraphics[width=0.45\textwidth]{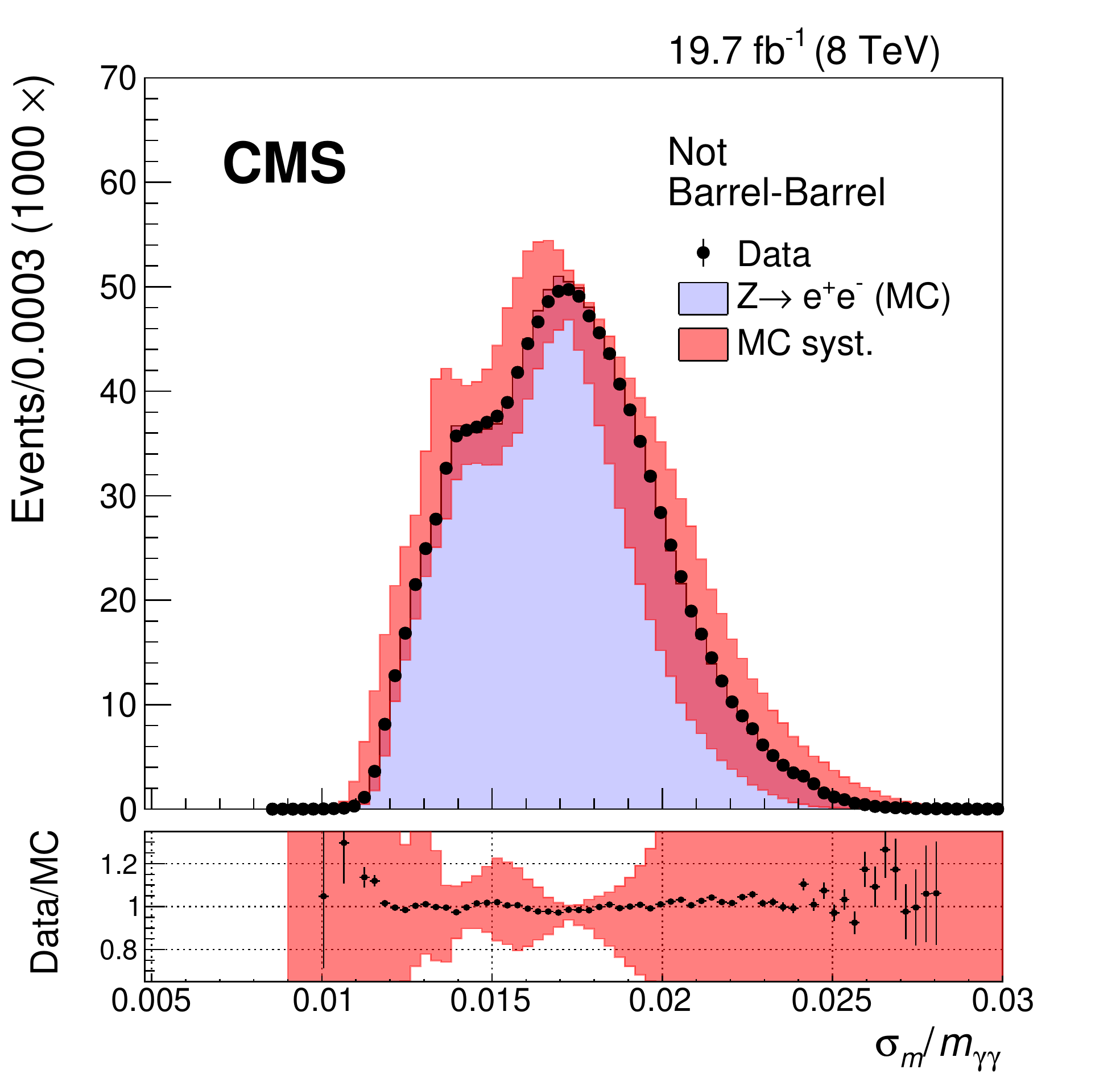}
   \caption{Mass resolution estimator $\sigma_m/\mgg$ after the decorrelation procedure, in \Zee events in data (dots) and simulated events (histogram) with their systematic uncertainties (shaded bands) for  barrel-barrel events (\cmsLeft),  all the other events (\cmsRight). The ratio of data to MC predictions are shown below each panel, and the error bars on each point represent the statistical uncertainties of the data.}
 \label{fig:EnergyResolutionZee}
\end{figure}

The value in the estimator of the mass resolution, after decorrelation, is used to categorize events. The $\sigma_m/\mgg$ distribution in $\Zee$ data shown in Fig.~\ref{fig:EnergyResolutionZee} indicates good agreement with the MC expectation over the whole range of $\sigma_m/\mgg$.
The double bump structure corresponds mainly to events where (Fig. 2a) both photons are in the central barrel ($\abs{\eta}<1.0$, $\sigma_m/\mgg < 1\%$), or at least one of the photons is in the outer barrel ($1.00<\abs{\eta}<1.44$, $\sigma_m/\mgg > 1\%$), and (Fig. 2b) one photon is in the barrel and one in the endcap, both having high values of the \RNINE\ ($\sigma_m/\mgg < 1.5\%$), and all the other events ($\sigma_m/\mgg > 1.5\%$).
The class boundaries in $\sigma_m/\mgg$ are optimized in MC samples simultaneously with the photon-identification working points desired to maximize signal significance.
The photon identification efficiency is about 95\% in each category of $\sigma_m/\mgg$. The category with best energy resolution has $\sigma_m/\mgg < 0.79\%$, and is composed of events with both selected photons in the central barrel.
Both the second ($0.79 < \sigma_m/\mgg < 1.28\%$) and third categories ($1.28 < \sigma_m/\mgg < 1.83\%$) have at least one photon in the outer barrel or in the endcap. Events with $\sigma_m/\mgg > 1.83\%$ do not provide a noticeable improvement in sensitivity and are therefore rejected.
Classifying events in these three categories improves the analysis sensitivity by nearly 10\%.

It was verified in the simulation that the classification of events according to $\sigma_m/\mgg$, after implementing the decorrelation procedure for $\sigma_E$, does not produce a distortion of the background distribution in \mgg.

\subsection{Jet identification}

Jets are reconstructed using particles identified by the PF algorithm, using the anti-\kt \cite{Cacciari:2008gp} algorithm with a distance parameter of 0.5.
Jet energy corrections account in particular for pileup, and are obtained from simulation. They are calibrated with in situ measurements using the energy balance studied in dijet and $\Pgg/\cPZ$+jet events~\cite{Chatrchyan:2011ds}.
The jet momentum scale is found to be within 5--10\% of the true jet momentum over the whole spectrum and detector acceptance.
The jet energy resolution is typically 15\% and 8\% at 10 and 100\GeV, respectively.
Mean resolutions of 10\% to 15\% are observed in the respective regions of $\abs{\eta}<0.5$ and $3<\abs{\eta}<5$.
Jets are selected if they fail the pileup identification criteria \cite{CMS-PAS-JME-13-005}, and have $\ptj>25$\GeV.
The minimum distance between photons and jets is required to be $\Delta R(\gamma,$j$)=\sqrt{\Delta\eta(\gamma,\mathrm{j})^2 + \Delta\phi(\gamma,\mathrm{j})^2}>0.5$, where $\Delta\eta(\gamma,$j$)$ and $\Delta\phi(\gamma,$j$)$ are the pseudorapidity and azimuthal angle differences between photons and jets, to minimize photon energy depositions into jets.

\subsection{Fiducial phase space and observables}

The generator-level fiducial volume is chosen to be close to that used in the selection of reconstructed events, and follows previous prescriptions: photons must have $\ptga/\mgg>1/3$ and $\ptgb/\mgg>1/4$, with both photons within $\abs{\eta}<2.5$. The photons have to be isolated at the generator level, with $\sum_{i} E_{\mathrm{T}i} < 10$\GeV, where $i$ runs over all the other generator-level particles in a cone $\Delta R<0.4$ around the photons.
This selection corresponds to a signal efficiency of 63\% in ggH, and almost 60\% considering other production mechanisms.
We measure the kinematic observables using two-photon criteria, as well as requiring at least one or two jets.

The transverse momentum $\ptgg$ and absolute value of the rapidity $\ygg$ of the Higgs boson are measured using the inclusive selection.
Both $\ptgg$ and $\ygg$ probe the production mechanism, while the photon helicity angle \costhetastar in the Collins--Soper frame \cite{CollinsSoper1977} of the diphoton system and the difference in azimuth $\deltaphigg$ between the two photons are related to properties of the decaying particles.

The number of jets $\Njets$, the transverse momentum $\ptja$ of the jet with largest (leading) $\pt$ in the event, and the rapidity difference between the Higgs boson and the leading jet $\yggj$ are defined after requiring at least one jet with $\pt>25$\GeV to lie within $\abs{\eta}<2.5$.
The difference in rapidities between the diphoton system and the leading jet provides a sensitive probe of any new contributions to the ggH process.

Requiring at least two jets with $\pt>25$\GeV and $\abs{\eta}<4.7$ provides the basis for defining the following observables: the dijet mass $\mjj$, the azimuthal angle difference of the two jets $\deltaphijj$, the difference in pseudorapidity $\deltaetajj$ between the leading and subleading jet, the Zeppenfeld \cite{Rainwater:1996ud} variable $\Zepp$, and the difference in azimuth between the Higgs boson and the dijet system $\deltaphiggjj$.
A requirement of $\abs{\eta^{\text{j}}}<2.5$ is applied in the single-jet selection, as this selection aims to probe primarily ggH process, while the two-jet selection has a requirement $\abs{\eta^{\text{j}}}<4.7$ since it is oriented toward VBF.

The bin boundaries for each kinematic observable are optimized to achieve similar relative statistical uncertainties in the expected cross section in each bin, namely 60\% for each bin in the inclusive observables, 70\% in the one-jet observables, and more than 100\% in the two-jet observables. The relative statistical uncertainty expected in the fiducial cross section is 30\%.

\section{Extraction of signal and unfolding of detector effects}
\label{sec:unfolding}

\newcommand{\Lk}{\ensuremath{{\cal L}}\xspace}
\newcommand{\Lt}{\ensuremath{\widetilde{\cal L}}\xspace}

The signal yield is extracted by fitting the $\mgg$ distribution using a signal model based on simulated events, and a background model determined in the fit to the data. The statistical methodology is similar to Ref.~\cite{HggLegacy}, and for each observable the fit is performed simultaneously in all the bins. The reconstructed yields are corrected for detector effects by including the response matrix in the fit.

\subsection{Models for signal}
\label{sec:sigmodel}

Signal models are constructed for each class of $\sigma_m/\mgg$ events,
and for each production mechanism, from a fit to the simulated \mgg distribution, after applying the corrections determined from comparisons of data and simulation for $\Zee$ (also checked with $\Zmmg$) events, for $\mH=120,125$, and $130\GeV$.
Mass distributions for the best and worst choices of diphoton vertex, corresponding to the highest and lowest scores in the vertexing BDT \cite{HggLegacy}, are fitted separately.
Good descriptions of the distributions can be achieved using sums of Gaussian functions, where the means are not required to be identical.
As many as four contributing Gaussian functions are used, although in most cases two or three provide an acceptable fit.
Models for intermediate values of \mH\ are obtained by linear interpolation of the fitted parameters.

\subsection{Statistical methodology}
\label{sec:bgmodel}

After implementing the above-described selection requirements on photon candidates, a simultaneous binned maximum likelihood fit is performed
to the diphoton invariant mass distributions in all the event classes
over the range $100<\mgg<180\GeV$ for each differential observable.
The test statistic chosen to measure signal and background contributions in data
is based on the profile likelihood ratio~\cite{LHC-HCG-Report, Chatrchyan:2012tx}.
Systematic uncertainties are incorporated into the analysis via nuisance parameters and treated according to the frequentist paradigm.

We use the same discrete profiling method to fit the background contribution~\cite{Dauncey:2014xga} as used in extracting the main \Hgg result \cite{HggLegacy}.
The background is evaluated by fitting the $\mgg$ distribution in data, without reference to MC simulation.
Thus the likelihood to be evaluated in a signal+background hypothesis is
\begin{equation}
\Lk(\mu_i)=\Lk(\text{data}|s_i(\mu_i,\mgg)+f_i(\mgg)),
\end{equation}
where $\mu_i$ is the signal strength (ratio of measured to expected yields) in the bin $i$ of a differential distribution that is varied in the fit, $s_i(\mu_i,\mgg)$ represents the model for signal, and $f_i(\mgg)$ the fitted background functions.

The choice of function used to fit the background in any
particular event class is included as a discrete nuisance parameter in the formulation used to extract the result.
Exponentials, power-law functions, polynomials in the Bernstein basis, and Laurent polynomials are used to represent $f(\mgg)$.
When fitting a signal+background hypothesis to the data, by minimizing the value of twice the negative logarithm of the likelihood, all functions in these families are tried, with a ``penalty" term added to account for the number of free parameters in the fit. The penalized likelihood function $\Lt_f$ for a single fixed background fitting function $f$ is defined through
\begin{equation}
-2\,\ln\Lt_f=-2\,\ln\Lk_f+N_{f},
\end{equation}
where $\Lk_f$ is the ``unpenalized'' likelihood function, and $N_{f}$ is the number of free parameters in $f$. The full set of $\mu_i$, denoted by $\vec{\mu}$, is determined by minimizing the likelihood ratio:
\begin{equation}
L(\vec{\mu})=-2\,\ln\frac{\Lt(\text{data}|\vec{\mu},\hat{\theta}_{\vec{\mu}},\hat{f}_{\vec{\mu}})}{\Lt(\text{data}|\hat{\vec{\mu}},\hat{\theta}_{\hat{\vec{\mu}}},\hat{f}_{\hat{\vec{\mu}}})},
\label{eq:q}
\end{equation}
where the numerator represents the maximum of $\Lt$ as a function of $\vec{\mu}$, achieved for the
best-fit values of the nuisance parameters $\theta_{\vec{\mu}}=\hat{\theta}_{\vec{\mu}}$, and a particular background function $f_{\vec{\mu}}=\hat{f}_{\vec{\mu}}$.
The denominator corresponds to the global maximum of $\Lt$, where $\vec{\mu}=\hat{\vec{\mu}}$, $\theta_{\vec{\mu}}=\hat{\theta}_{\hat{\vec{\mu}}}$, and $f_{\vec{\mu}}=\hat{f}_{\hat{\vec{\mu}}}$.
In each family, the number of degrees of freedom (number of exponentials, number of terms in the series, degree of the polynomial, etc.)
is increased until no significant improvement
($p\text{ value}<0.05$ obtained from the F-distribution~\cite{F-dist})
occurs in the likelihood between N+1 and N degrees of freedom for the fit to data.

For a given observable, the fit is performed simultaneously over all the bins, and
the nuisance parameters are profiled in the fit.
The signal mass is also considered a nuisance parameter and profiled for each observable. This choice is made to avoid using the same data twice, first to measure the signal mass, then to measure the kinematic distribution at this mass. As a consequence the differential cross section for each observable is evaluated at slightly different best fit values of the signal mass (see appendix A).
As an example, Fig.~\ref{fig:SignalBackgroundFit} shows the sum of the fit to the events of the three $\sigma_{m}/\mgg$ classes in the fiducial phase space measurement, under the signal+background hypothesis (S+B), weighted by S/(S+B) separately in each category.

\begin{figure}[htb]
 \centering
   \includegraphics[width=0.5\textwidth]{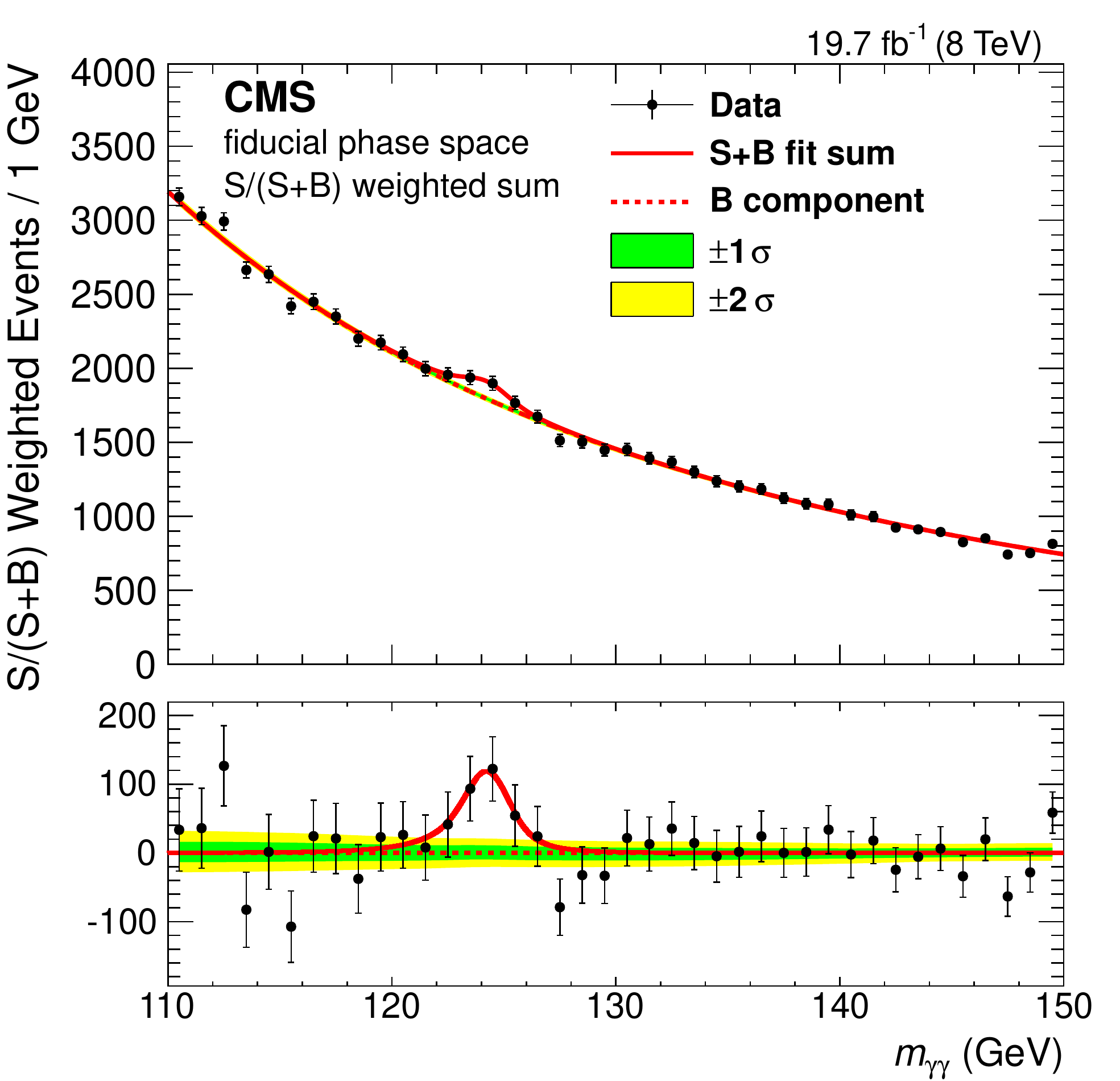}
   \caption{Sum of the signal+background (S+B) model fits to the events of the three $\sigma_{m}/\mgg$ classes in the fiducial phase space measurement, weighted by S/(S+B) separately in each category, together with the data binned as a function of $\mgg$. The 1 and 2 standard deviation bands of uncertainty (labeled as 1$\sigma$ and 2$\sigma$) shown for the background component include the uncertainty due to the choice of function and the uncertainty in the fitted parameters. The bottom panel shows the result after subtracting the background component.}
  \label{fig:SignalBackgroundFit}

\end{figure}

The uncertainty on the expected signal strength of the fiducial cross section is $\sigma_{\hat{\mu}} = 0.32$ for the present analysis, compared with the uncertainty on the expected signal strength obtained with the reference analysis of $\sigma_{\hat{\mu}} = 0.26$ using 8\TeV data. The reference analysis \cite{HggLegacy} classifies events using exclusive categories dedicated to measuring signal production in associated mechanisms, while the present analysis is performed inclusively.

\subsection{Unfolding detector effects}

The measurement is performed simultaneously in all bins, together with the unfolding of the detector effects to the particle level.
The same kind of procedure was used to extract the signal strength in untagged and dijet-tagged categories to  measure the couplings of the Higgs boson to vector bosons and fermions \cite{HggLegacy}. This procedure uses asymmetric uncertainties in the full likelihood instead of being limited to Gaussian uncertainties computed with a covariance matrix.

The unfolding of reconstructed distributions is based on models of response matrices $K_{ij}$ in the three $\sigma_m/\mgg$ categories for each observable to be measured.
The $K_{ij}$ are constructed in the simulation to give the probability of measuring a reconstructed event in bin $j$, given that it was generated in bin $i$.
The models for contributions of signal are contained in each element of the response matrix, which is mostly diagonal, but can have non-negligible bin-by-bin migration resulting in off-diagonal contributions.

The unfolding is performed using a maximum likelihood technique, adapted to combine the measurement in different categories. Regularization is not used, so as not to shift the best-fit value while artificially decreasing the uncertainties. This leads to minimizing the following conditional log-likelihood expression:
\begin{equation}
\mathcal{F} (\vec{\mu} ) = -2 \sum_j \log \mathcal{L} \Big (\sum_i K_{ij} \mu_i N^{\text{gen}}_{i}|N^{\text{reco}}_{j} \Big)
\end{equation}
where $\mathcal{L}$ is the log-likelihood expression in Eq.~(2), $\mu_i N^{\text{gen}}_{i}$ is the unknown unfolded particle-level distribution, $\mu_i$ is the unknown signal strength at particle level, $N^{\text{gen}}_{i}$ is the particle-level distribution in the simulated kinematic observable, and $N^{\text{reco}}_{j}$ is the number of events in each bin of the measured distribution. The indices $i$ refer to particle-level bins while $j$ refer to reconstructed-level bins in the three $\sigma_m/\mgg$ categories.

The expected reconstructed signal at detector level is given by the vector $J_{j}=K_{ij} N^{\text{gen}}_{i}$, where each entry of $K_{ij}$ corresponds therefore to a set of signal models, computed by interpolating the matrix between the generated mass points, weighted by the signal efficiency interpolated at the same mass.
The matrix is a function of $\mH$, the model for the generated bin $i$ and the reconstructed bin and category $j$, as well as all the nuisance parameters.
Events falling out of the acceptance are taken into account by forming an extra bin.

The maximum likelihood fit is performed simultaneously for the diphoton background and the signal strength in each generator bin, as described in Section~\ref{sec:bgmodel}.
The out-of-acceptance bin is left fixed in the fit, because there are only $N$ bins at the reconstructed level, where the detector is able to perform the measurement, and it is therefore not possible to determine an extra unknown at the particle level ($N+1$ unknowns), outside of the detector acceptance.
To restore the correct cross section normalization in the fiducial region, the generator distributions for the variables of interest ($N^{\text{gen}}_{i}$ at the fitted mass point $\mH$) are multiplied by the measured set of signal strengths ($\mu_i$).

The enhancement of the statistical uncertainties due to the presence of off-diagonal elements in the response matrix is small for the inclusive observables, while it is non-negligible for the one-jet or two-jet observables.
A negative number of events is measured in only one bin of the rapidity difference between the Higgs boson and the leading jet, in the last bin of the dijet mass distribution, and in two bins of the azimuthal difference between the Higgs boson and the dijet system. In each case, within the uncertainties, the result is compatible with zero.

The model dependence introduced by the unfolding is checked using the following procedure.
The same model used for the SM Higgs boson is kept in the unfolding matrix, leaving the expected yield for ggH unchanged. The expected number of signal events arising from associated production mechanisms is altered by 50\% in the fit.
This change introduces a redistribution of the events in the $\sigma_m/\mgg$ categories relative to the nominal analysis.
The change in the fiducial cross section is less than 5\% of its statistical uncertainty. In general, the impact in each bin of the measured differential distributions can be up to 5\% and 10\% of the statistical uncertainty in the respective bins of the inclusive and jet observables.

\section{Systematic uncertainties}
\label{sec:systematics}

Systematic uncertainties listed in this section are included in the likelihood as nuisance parameters and are profiled during the minimization. Unless specified to the contrary, the sources of uncertainty refer to the individual quantity studied, and not to the final yield. The precision of the present measurement is however dominated by statistical uncertainties.

The sources of uncertainty assigned to all events can be summarised as follows.
The uncertainty in the integrated luminosity is estimated as described in Ref.~\cite{CMS-PAS-LUM-13-001},
and amounts to 2.6\% of the signal yield in the data.
The uncertainty in the vertex-finding efficiency is taken from the
difference observed between data and simulation
in $\Zmm$ events, following removal of the muon tracks to mimic a diphoton event.
A 1\% uncertainty is added to account for the activity from charged-particle tracks in signal, estimated by changing the underlying event tunes in ggH events; another uncertainty of 0.2\% accounts for the uncertainty in the \mgg distribution of signal events.
The uncertainty in the trigger efficiency is extracted from $\Zee$ events using a ``tag-and-probe" technique \cite{TagAndProbe}.
A rescaling in the \RNINE~distribution is used to take into account the difference between electrons and photons, for a total uncertainty of 1\% assigned for this source.

The following correspond to systematic uncertainties related to individual photons.
The uncertainty in the energy scale of photons
is assessed using simulated samples in which the amount of tracker material is increased uniformly by 10\% in the central barrel, where the material is known with best precision, and 20\% out of this region. These values were chosen as upper limits on the additional material, as derived from the data. The resulting uncertainty in the photon energy ranges from 0.03\% in the central ECAL barrel up to 0.3\% in the outer endcap.
Additional uncertainties of 0.015\% are due to the modelling of the fraction of scintillation light reaching the photodetector, and from nonuniformities in the radiation-induced loss of transparency of the crystals.
A small uncertainty of 0.05\% is added to account for modelling of electromagnetic-showers in $\GEANT4$~version~9.4.p03.

Possible differences between MC simulation and data in the extrapolation of shower energies typical of electrons from $\Zee$ decays
to those typical of photons from $\Hgg$ decays, have been investigated with $\Zee$ and $\Wenu$ data.
The effect of differential nonlinearity in the measurement of photon energies has an effect of up to 0.1\% on the diphoton mass for $\mgg \approx 125\GeV$.

The energy scale and resolution in data are measured with electrons from
$\Zee$ decays. Systematic uncertainties in the method are estimated as a function of $\abs{\eta}$ and $\RNINE$.
The uncertainties range from 0.05\% for unconverted photons in the central ECAL barrel, to 0.1\% for converted photons in the outer endcaps of the ECAL.
Finally, there is an overall uncertainty that accounts for possible mismodelling of the $\Zee$ line shape in simulation.

The uncertainties in the BDT discriminant for photon identification and in the estimate of photon energy resolution are discussed together since they are studied in the same way.
The dominant underlying cause of the observed differences between data and simulation
is the simulation of the energy distribution in the shower.
The combined contribution of the uncertainties in these two quantities dominates
the experimental contribution to the systematic uncertainty in signal strength.
The agreement between data and simulation is examined when photon candidates are
electron showers reconstructed as photons in $\Zee$ events, photons in $\Zmmg$ events, and
leading photons in preselected diphoton events where $\mgg>160\GeV$.
A change of $\pm0.01$ in the value of the photon identification discriminant, together with an uncertainty in the estimated photon energy resolution parameterized, respectively as a rescaling of the resolution estimate by $\pm5\%$ and $\pm10\%$ about its nominal value in the barrel and in the endcap, fully cover the differences observed in all three of these data samples.

The uncertainty in the preselection efficiency is taken as the uncertainty in the data/MC scale factors measured using $\Zee$ events with a tag-and-probe technique.

Jet observables are affected by systematic uncertainties arising from jet identification, jet energy scale, and resolution.
For the jet observables, a systematic uncertainty of less than 1\% in the impact of the algorithm used to reject jets from pileup is neglected.
For jets within $\abs{\eta}<2.5$, the energy scale uncertainty is $\approx$3\% at 30\GeV, and decreases quickly as a function of the increasing jet $\pt$.
The impact of this uncertainty in the cross section is 1--5\%, increasing with the number of jets.
Dijet observables for forward jets (up to $\abs{\eta}<4.7$) have the worst energy resolution and a scale uncertainty of $\approx$4.5\% at 30\GeV.
The impact of these uncertainties on the cross section ranges from 5 to 11\%, increasing in kinematical regions of observables where jet $\eta$ is large.
Small contributions from corrections in jet energy resolution have an impact of less than 1\% and are neglected.

\section{Comparison of data with theory}
\label{sec:theory}

\subsection{Theoretical predictions}

{\tolerance=800
The unfolded data are compared with the \HRES~\cite{HRes1,HRes2}, $\POWHEG$~\cite{powheg1,powheg2,powheg3,powheg-ggH}, \POWHEGMINLO~\cite{PowhegMinlo}, and \MADGRAPHAMCATNLO~\cite{aMC@NLO,mgamcatnlo} MC generators for ggH production.
\par}

The \HRES parton-level generator corresponds to next-to-next-to-leading order (NNLO) accuracy in perturbative QCD, with next-to-next-to-leading logarithm soft-gluon resummation; \HRES v.2.3 assumes finite bottom and top quark masses, using respective values of $m_b=4.75$\GeV and $m_t=175$\GeV. The renormalization scale $\mu_R$ and factorization scale $\mu_F$ are set to $\mH=125$\GeV, while the resummation scale is set to $\mH/2$. The MSTW2008NNLO \cite{MSTW2008} PDF is used for the central value, and its 68\% confidence level eigenvectors for computing the uncertainty (following the LHAPDF \cite{LHAPDF} recipe). The dependence on scale is evaluated by changing independently both the renormalization and factorization scale up and down by a factor of two around the central value $\mH$, and not considering simultaneous changes such as $\mu_R=\mH/2$ and $\mu_F=2 \mH$.
Because \HRES cannot be interfaced to a parton-shower program, no isolation is applied at the partonic level. A nonperturbative correction must be applied to the distributions to correct for the efficiency loss due to isolation requirements in the presence of parton shower and underlying event.
The nonperturbative correction is evaluated from the mean of the isolation efficiencies computed with $\POWHEG$ and \MADGRAPHAMCATNLO (as described below), estimated to be 3.1\% (up to 5\% in some bins).
The uncertainty is taken as half of the envelope, which is between 0.5\% to 5\%, depending on the kinematical region.

The $\POWHEG$ parton-level generator implements NLO calculations \cite{powheg1} interfaced to parton shower programs.
Samples of events with a Higgs boson with $\mH=125$\GeV produced via ggH, assuming an infinite top quark mass,
are generated and hadronized with \PYTHIA 6.4.
A sample of events with a Higgs boson produced in association with just a single jet, called $\POWHEG$ HJ, is also generated with \POWHEGMINLO~\cite{PowhegMinlo}. This sample has NLO accuracy for 0-jet and 1-jet production, while it is only leading-order (LO) for 2-jet final states.
Both samples set the damping factor $\textit{hfact}$ in $\POWHEG$ at 100\GeV to reproduce the predicted $\pt$ distribution of the Higgs boson from \HqT \cite{HqT1, HqT2}.
This factor minimizes emission of extra jets beyond those in the matrix element in the limit of large $\pt$,
and enhances contribution from the $\POWHEG$ Sudakov form factor as $\pt$ approaches 0.
The CT10 PDF and \PYTHIA6 tune Z2* are used in the calculation. Theoretical uncertainties are computed in the same way as described for \HRES.

The \MADGRAPHAMCATNLO matrix element generator is capable of generating LO and NLO processes \cite{mgamcatnlo}.
The ggH process is generated using the NLO Higgs characterization model \cite{higgsCharacterization}, with effective coupling of the Higgs boson to gluons in the infinite top quark mass limit.
Gluon fusion is generated with 0, 1, or 2 additional jets at NLO in the Born matrix element, and combined using FxFx merging~\cite{fxfxmerging}.
Samples are generated using the CT10 PDF, and showered using \PYTHIA 8.185 \cite{pythia8} with the 4C tune \cite{Corke:2010yf}. A nominal merging scale of 30\GeV is used for the additional jet multiplicities.
The effect of changing the merging scale from 20 to 60\GeV is very small compared to uncertainties in scale and in the choice of PDF.
Uncertainties from renormalization and factorization scales are evaluated in the same way as with \HRES and $\POWHEG$.

{\tolerance=1600
Theoretical predictions for associated production mechanisms are computed with the following generators. $\POWHEG$ interfaced with \PYTHIA6 is used for VBF, while standalone \PYTHIA6 is used for VH and \ttH.
In the following, the notation XH refers to the sum of VBF, VH and \ttH predictions for these generators. Each of the ggH predictions for \HRES, $\POWHEG$, \MADGRAPHAMCATNLO, \POWHEGMINLO, and XH processes are normalized to the total cross sections from Ref.~\cite{LHCHiggsCrossSectionWorkingGroup:2013}.
\par}

Along with the SM predictions, the following alternative models are considered. Spin \twomp minimal model (graviton-like) initiated through two production mechanisms: ggH and $\qqbar$ annihiliation, based on the \textsc{jhugen} generator \cite{Gao:2010qx, Bolognesi:2012mm} and normalized to the total SM cross section. The main changes relative to the SM are expected in the inclusive Collins--Soper \costhetastar angular distribution in the $\gamma\gamma$ rest frame, which provides maximum information on the spin of the $\gamma\gamma$ system. The two spin-2 samples are compared to the data in the \costhetastar observable.
 Anomalous couplings parametrized with the $O_W$ dimension-6 operator in linearly realized effective field theory \cite{Corbett:2013pja} are also considered, and implemented through the Universal FeynRules Output \cite{Degrande:2011ua} in \MADGRAPH 5. The $O_W$ operator is related to the anomalous triple gauge coupling parameter $\Delta g_1^Z$ \cite{Altarelli:1996gh}. The values of the Wilson coefficients are $F_W = -5 \times 10^{-5}\GeV^{-4}$ and $F_W = +5 \times 10^{-5}\GeV^{-4}$, both corresponding to $\Delta g_1^Z = 0.21$, a value approximately five times the size of the limits set by LHC diboson measurements \cite{Chatrchyan:2013yaa,Aad:2012twa,ATLAS:2012mec,Khachatryan:2015sga}. Both values modify the kinematic distributions of the Higgs boson in the VBF process toward larger $\pt$, and the $F_W = -5 \times 10^{-5}\GeV^{-4}$ value also increases the VBF cross section by approximately a factor of 3.

{\tolerance=600
All the predictions are generated at $\mH=125$\GeV. For each observable, a correction factor is applied in each bin of the differential cross section to correct for the mass difference of the generated sample relative to the measured $\mH$ in data. The correction is computed with \POWHEG{}+\PYTHIA8 for both the ggH and VBF processes, and with \PYTHIA6 for VH and \ttH processes. It amounts to less than 1\% for all bins, and integrates to a 0.8\% effect in the fiducial cross section.
\par}

\subsection{Results}

The fiducial cross section, inclusive in the number of jets, is measured to be:
\begin{equation*}
\sigma_{\text{obs}} = 32^{+10}_{-10}\stat^{+3}_{-3}\syst\unit{fb},
\end{equation*}
where the uncertainties reflect statistical and systematic contributions added in quadrature. This can be compared with the following SM predictions:
\begin{align*}
\sigma_{\HRES+\mathrm{XH}} &=  31 ^{+4}_{-3}\unit{fb},\\
\sigma_{\POWHEG+\mathrm{XH}} &= 32 ^{+6}_{-5}\unit{fb},\\
\sigma_{\MADGRAPHAMCATNLO+\mathrm{XH}} &= 30 ^{+6}_{-5}\unit{fb}.
\end{align*}

Uncertainties in the predicted cross sections include contributions from renormalization and factorization scales, choice of PDF and branching fraction. The \HRES also includes uncertainties from nonperturbative corrections.

The observed fiducial cross section agrees with the predicted values. The measurement precision is dominated by statistical uncertainties. The relative systematic uncertainty of 9\% is almost negligible relative to the statistical uncertainty of 30\%. The experimental uncertainty is larger than the theoretical one by a factor up to about two. The ratio of the measured cross section to the predictions for $\POWHEG$+$\mathrm{XH}$ is in good agreement with the signal strength observed in Ref.~\cite{HggLegacy}.

\begin{figure*}[htbp]
 \centering
   \includegraphics[width=0.45\textwidth,]{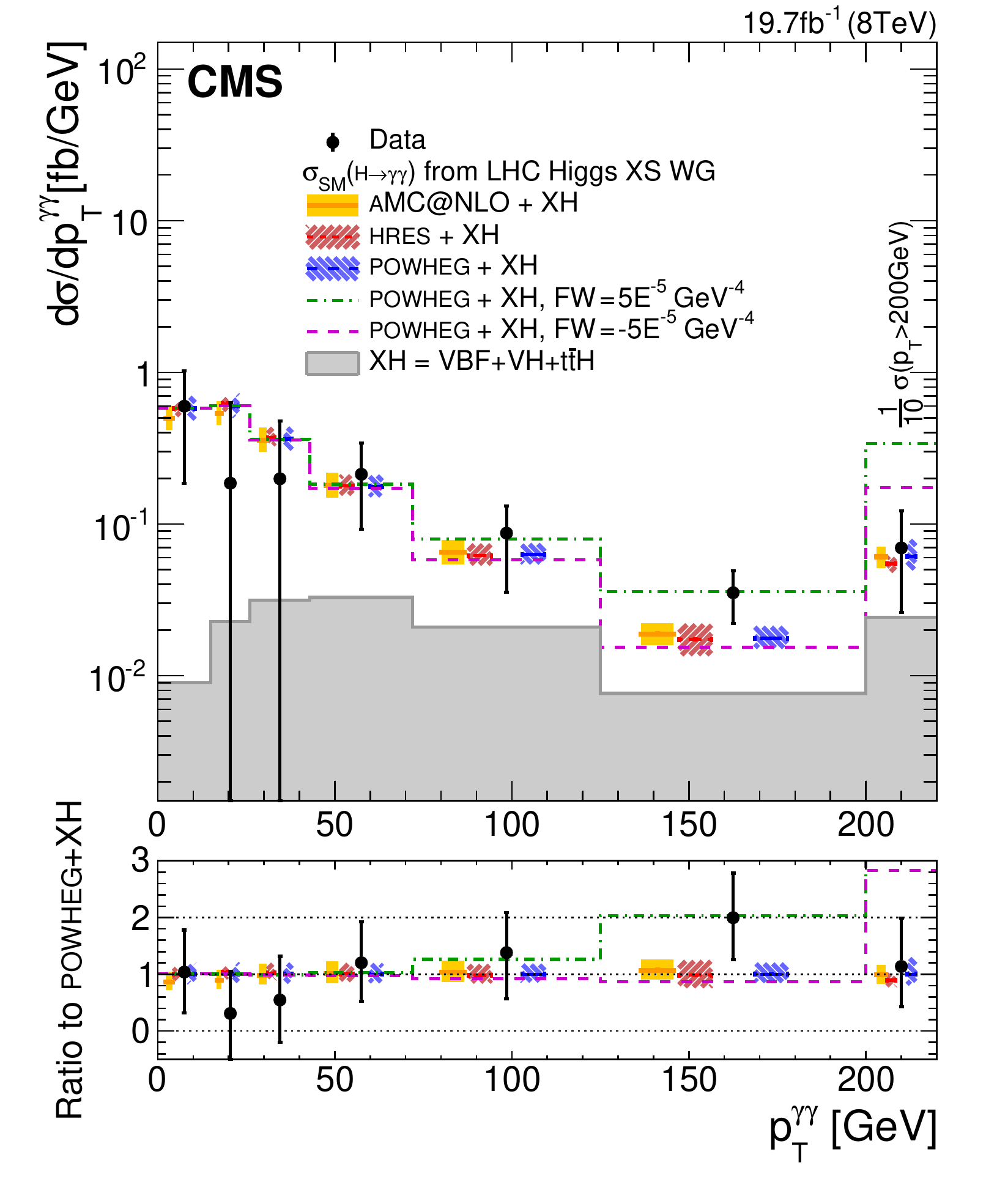}
   \includegraphics[width=0.45\textwidth,]{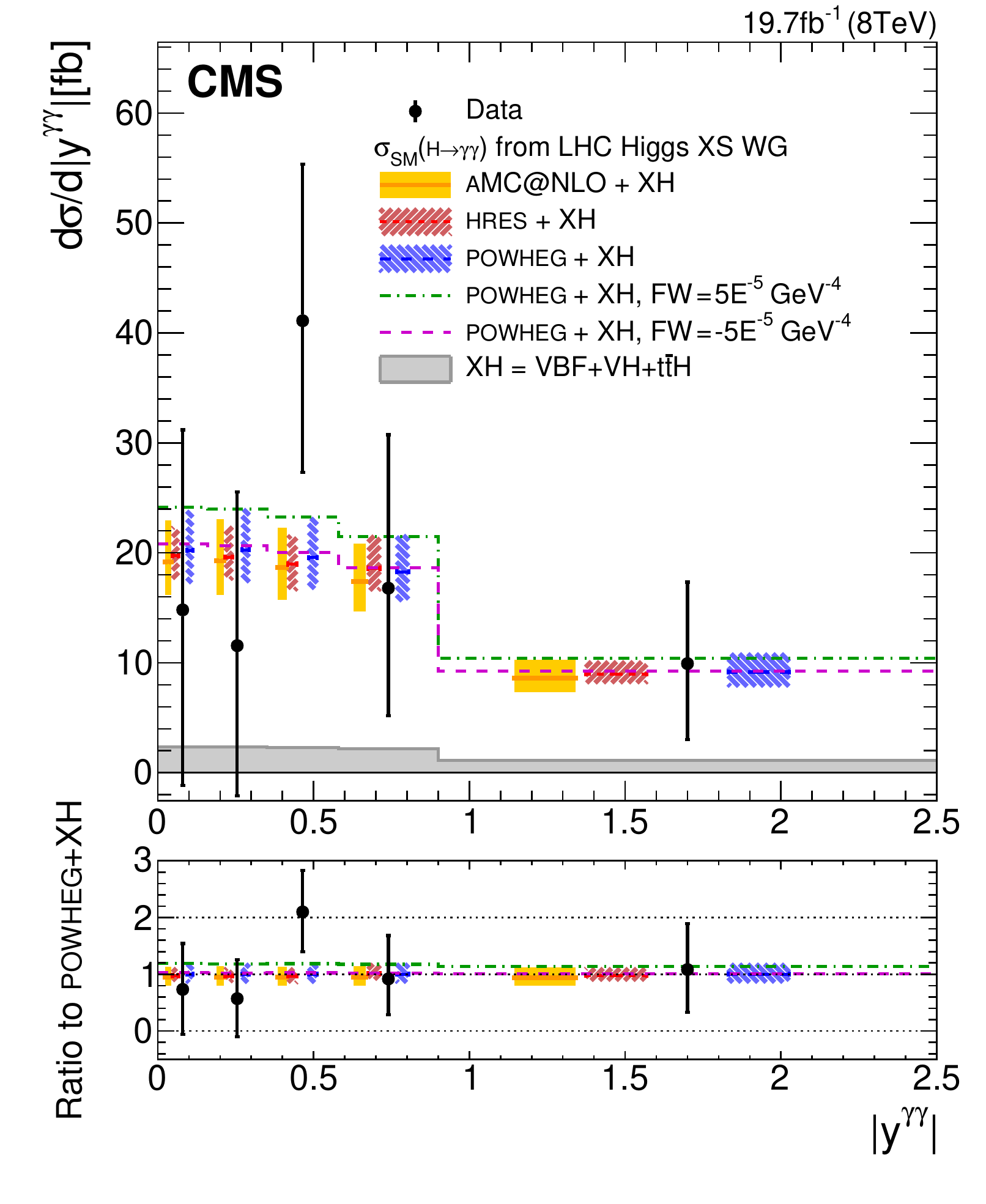}
   \includegraphics[width=0.45\textwidth,]{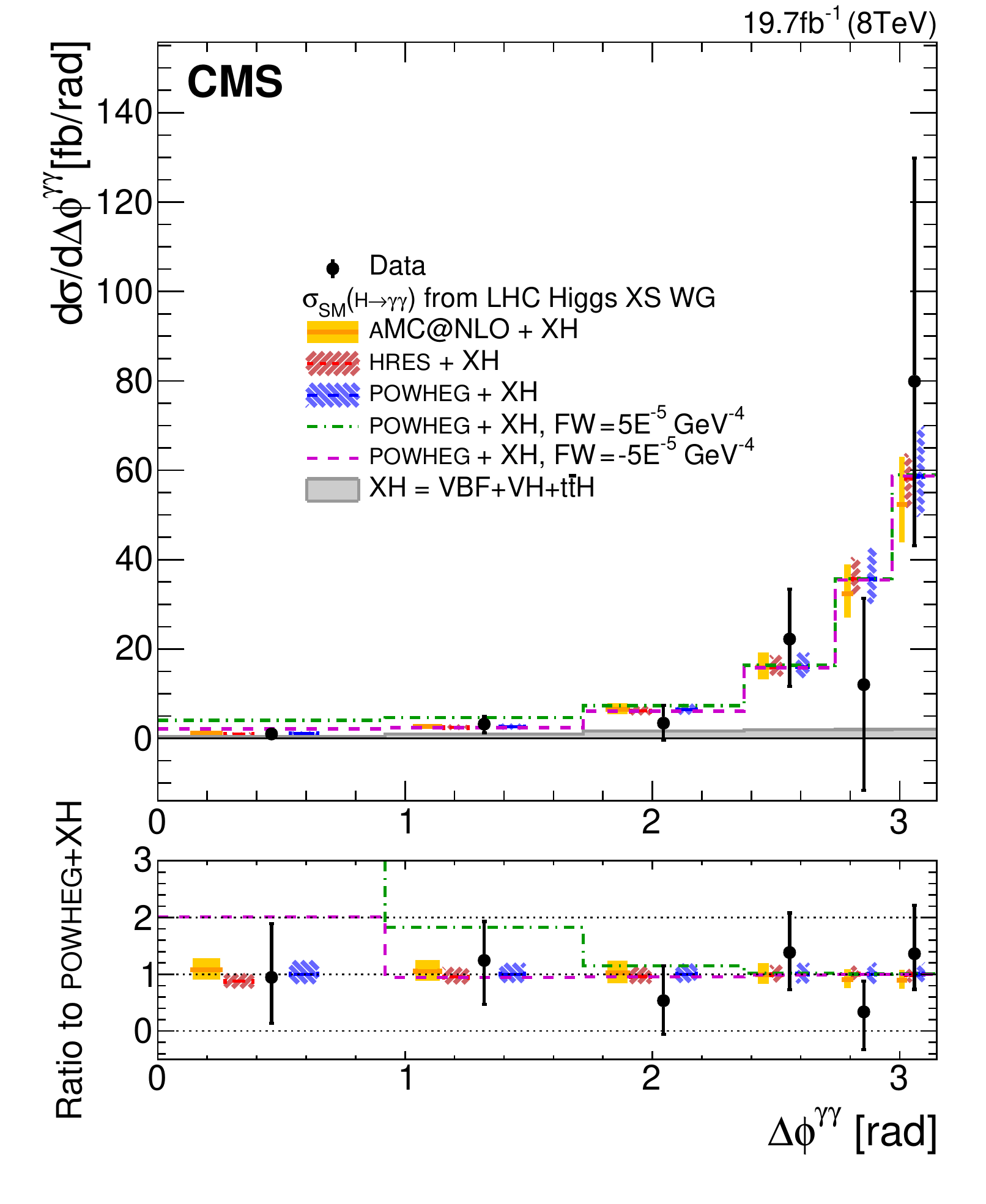}
   \includegraphics[width=0.45\textwidth,]{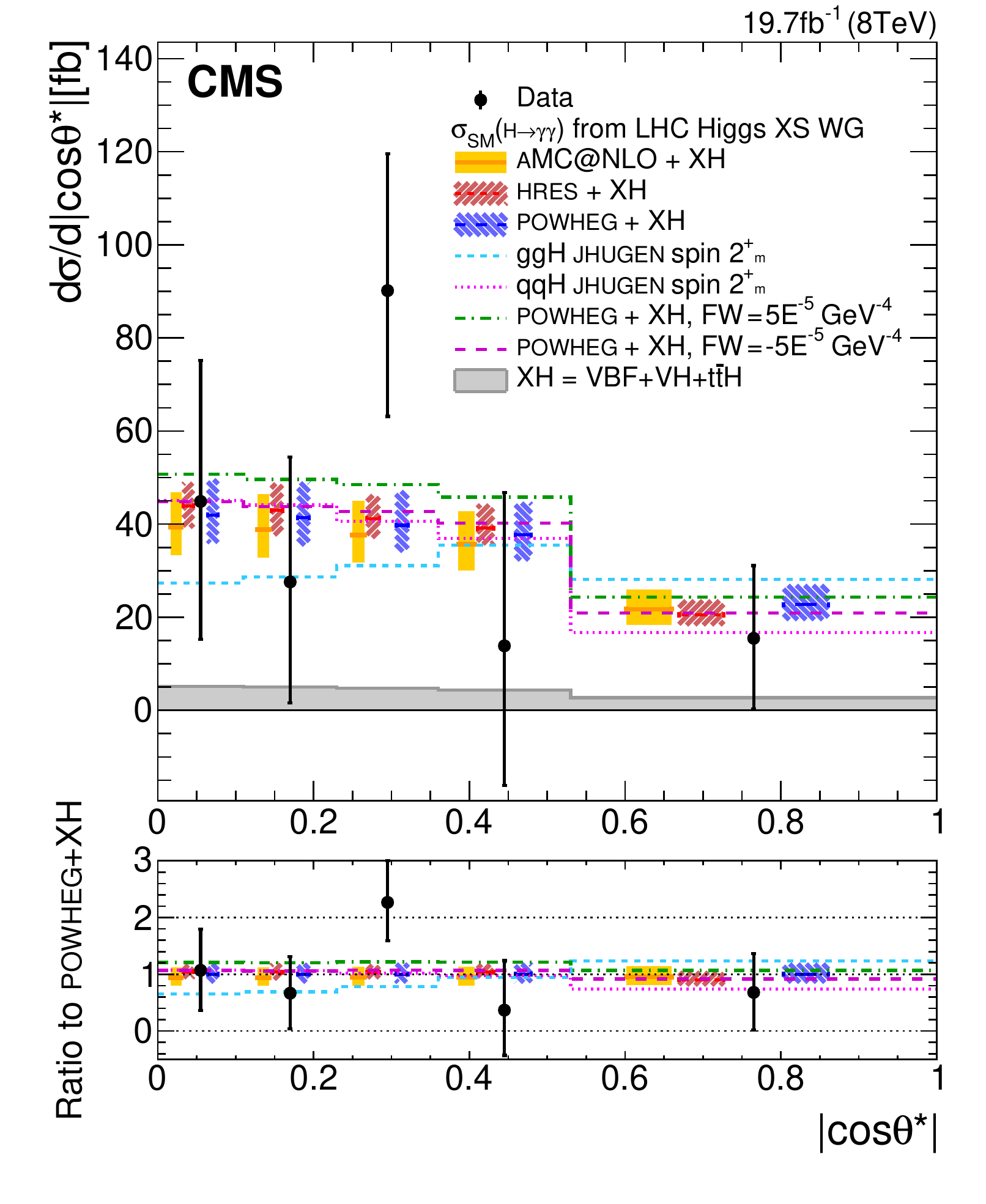}
   \caption{The $\Hgg$ differential cross section for inclusive events as a function of (upper left) $\ptgg$, (upper right) $\ygg$, (lower left) $\deltaphigg$, and (lower right) $\abs{\costhetastar}$. All the SM contributions are normalized to their cross section from Ref.~\cite{LHCHiggsCrossSectionWorkingGroup:2013}. Theoretical uncertainties in the renormalization and factorization scales, PDF, and branching fraction are added in quadrature. The error bars on data points reflect both statistical and systematic uncertainties. The last bin of $\ptgg$ distribution sums the events above 200\GeV. For each graph, the bottom panel shows the ratio of data to theoretical predictions from the $\POWHEG$ generator.}
  \label{fig:DataTheoryComp_Inclusive}
\end{figure*}

\begin{figure*}[htbp]
 \centering
   \includegraphics[width=0.45\textwidth,]{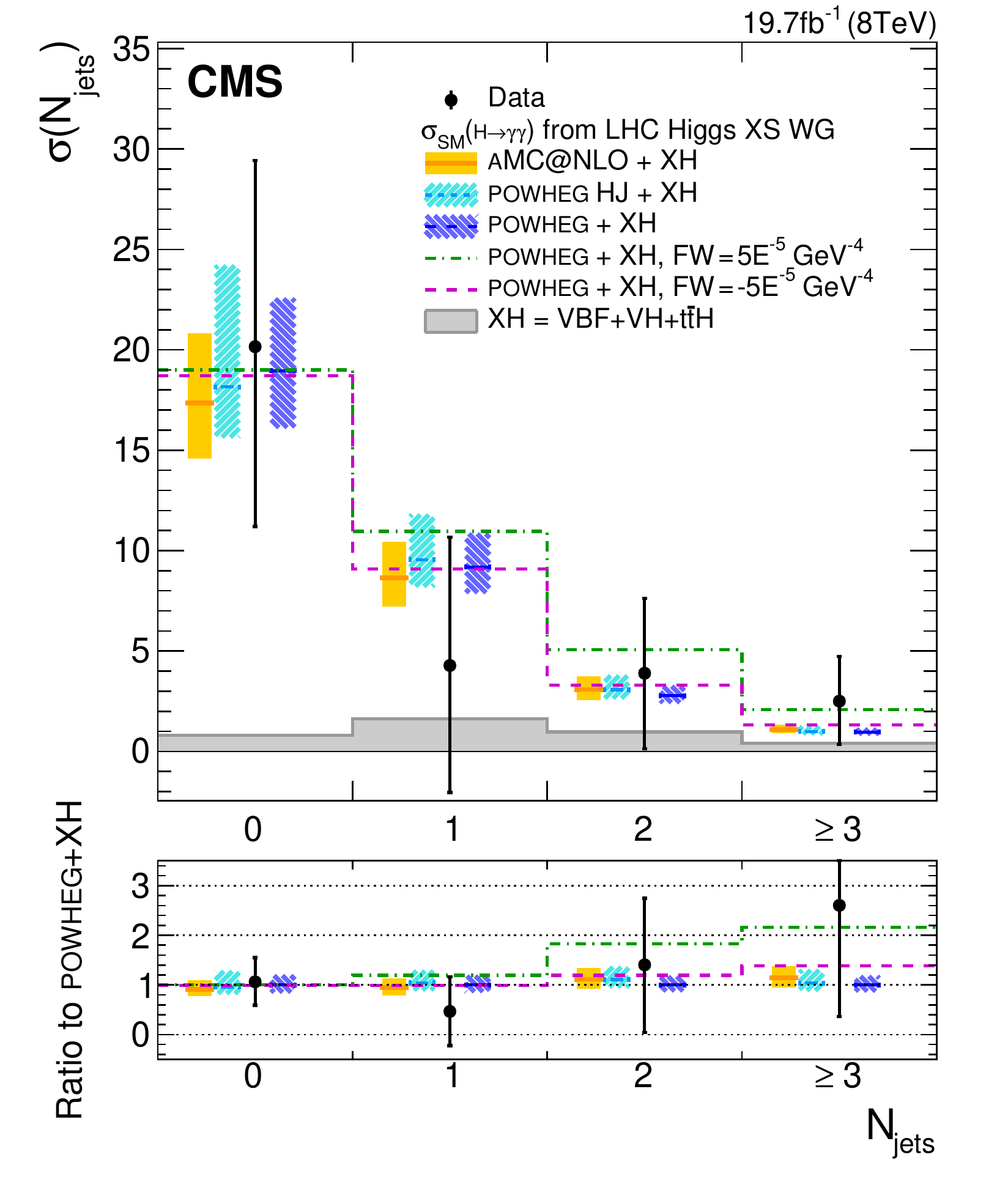}
   \includegraphics[width=0.45\textwidth,]{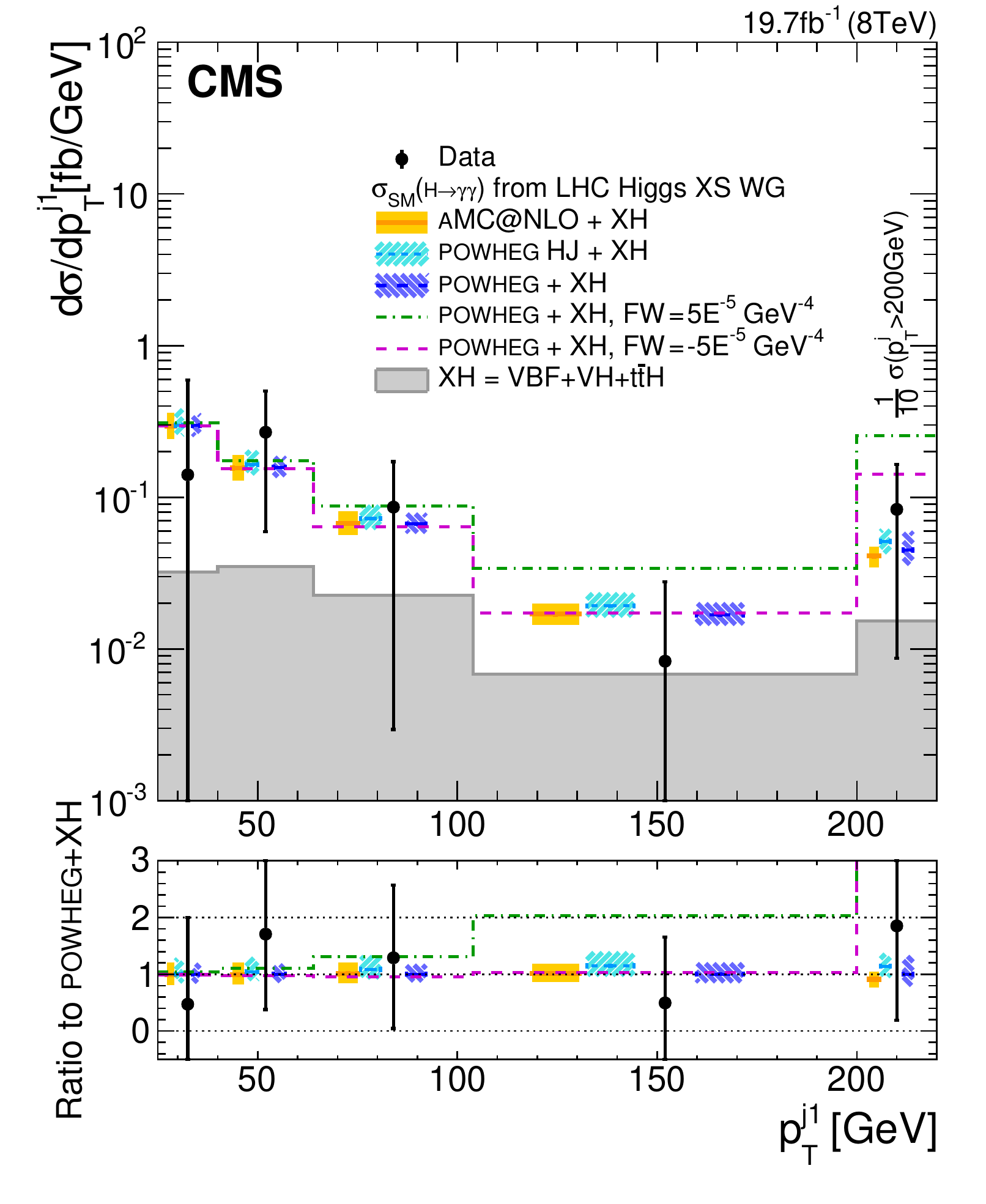}
   \includegraphics[width=0.45\textwidth,]{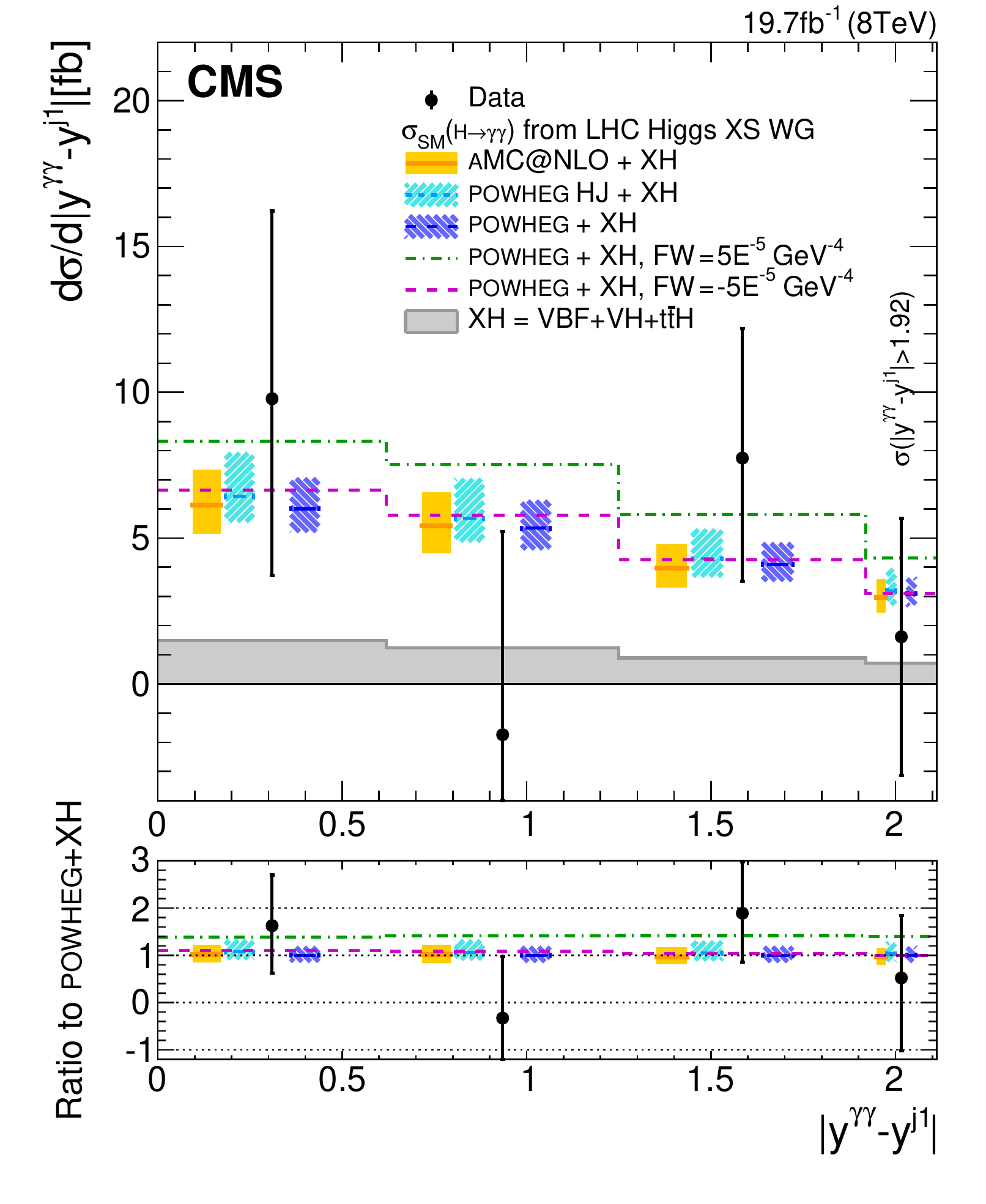}
   \includegraphics[width=0.45\textwidth,]{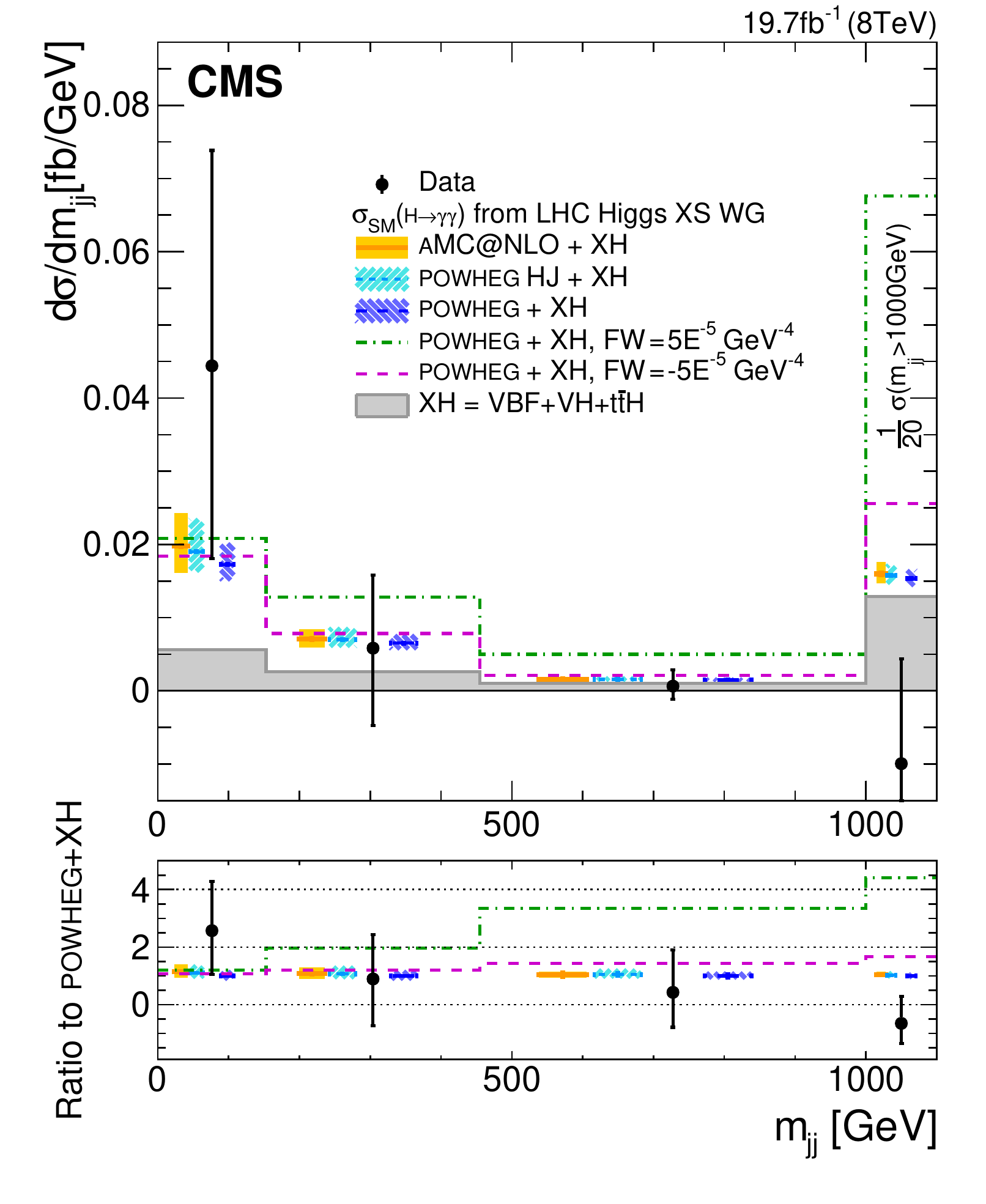}
   \caption{The $\Hgg$ differential cross section for H+jets events as a function of (upper left) $\Njets$, (upper right) $\ptja$, (lower left) $\yggj$, with jets within $\abs{\eta}<2.5$, and (lower right) $\mjj$ with jets within $\abs{\eta}<4.7$. All the SM contributions are normalized to their cross section from Ref.~\cite{LHCHiggsCrossSectionWorkingGroup:2013}. Theoretical uncertainties in the renormalization and factorization scales, PDF, and branching fraction are added in quadrature. The error bars on data points reflect both statistical and systematic uncertainties. In each distribution, the last bin corresponds to the sum over the events beyond the bins shown in the figure. For each graph, the bottom panel shows the ratio of data to theoretical predictions from the \POWHEG generator.}
  \label{fig:DataTheoryComp_1j_Eta25}
\end{figure*}

\begin{figure*}[htbp]
 \centering
   \includegraphics[width=0.45\textwidth,]{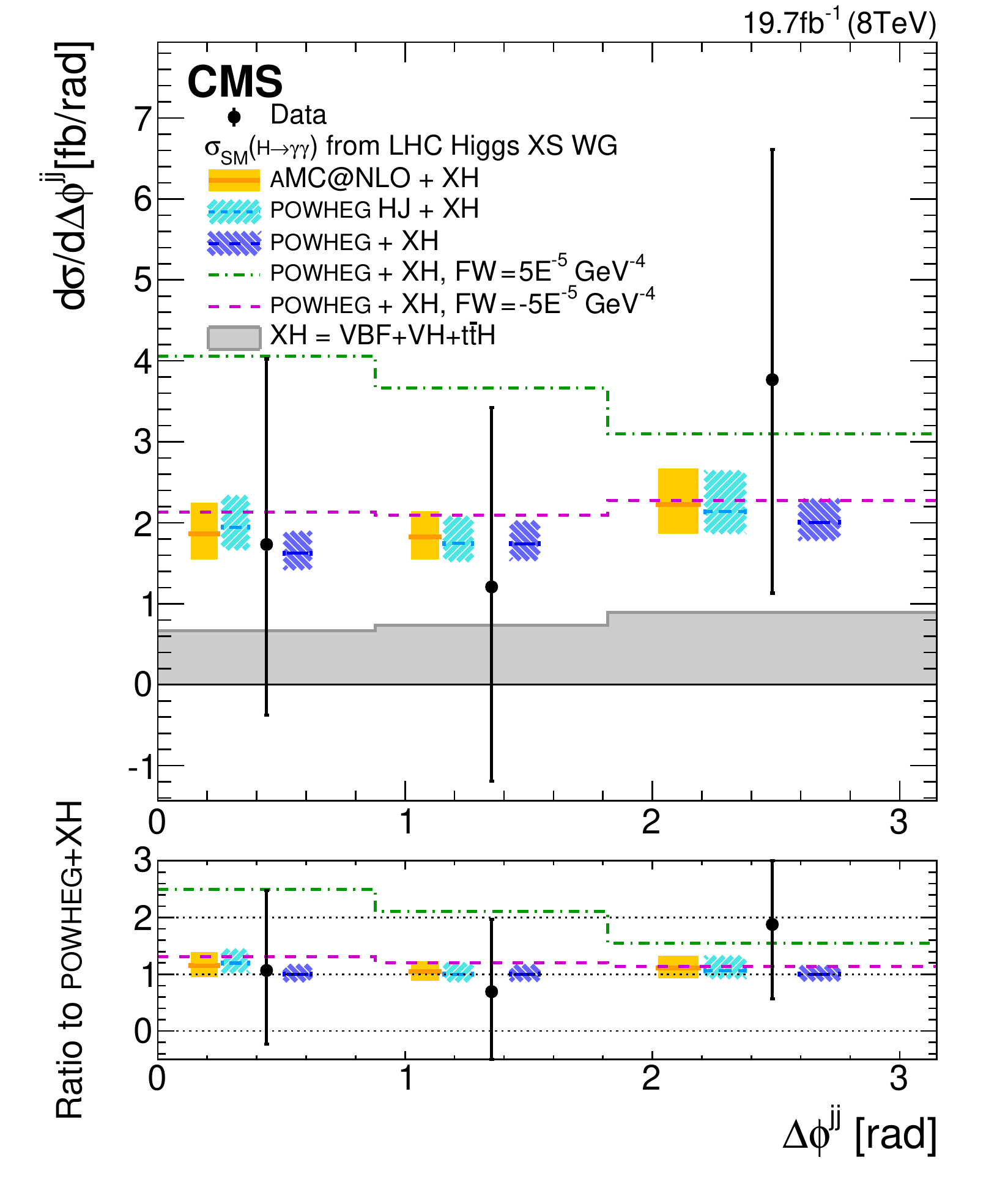}
   \includegraphics[width=0.45\textwidth,]{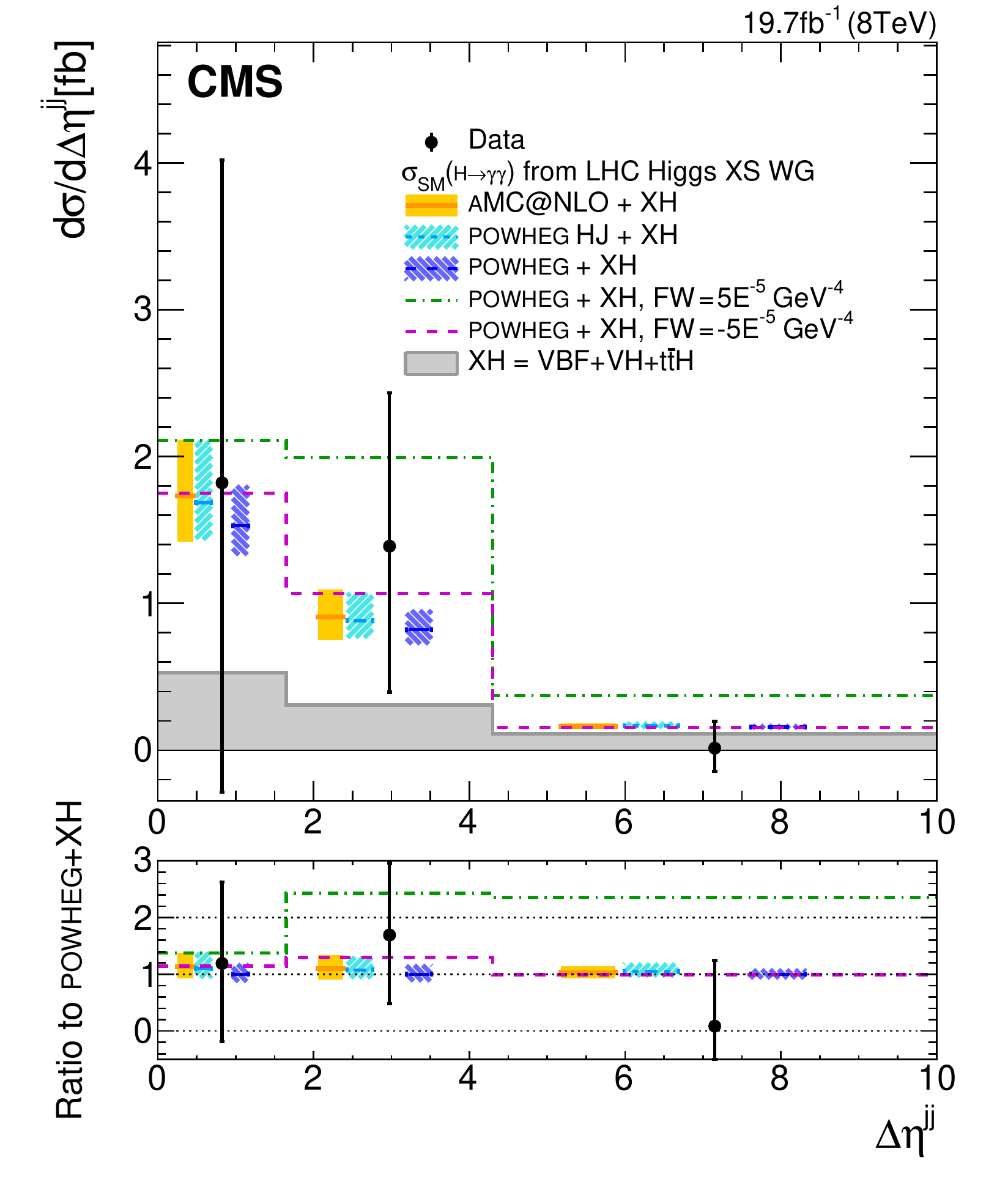}
   \includegraphics[width=0.45\textwidth,]{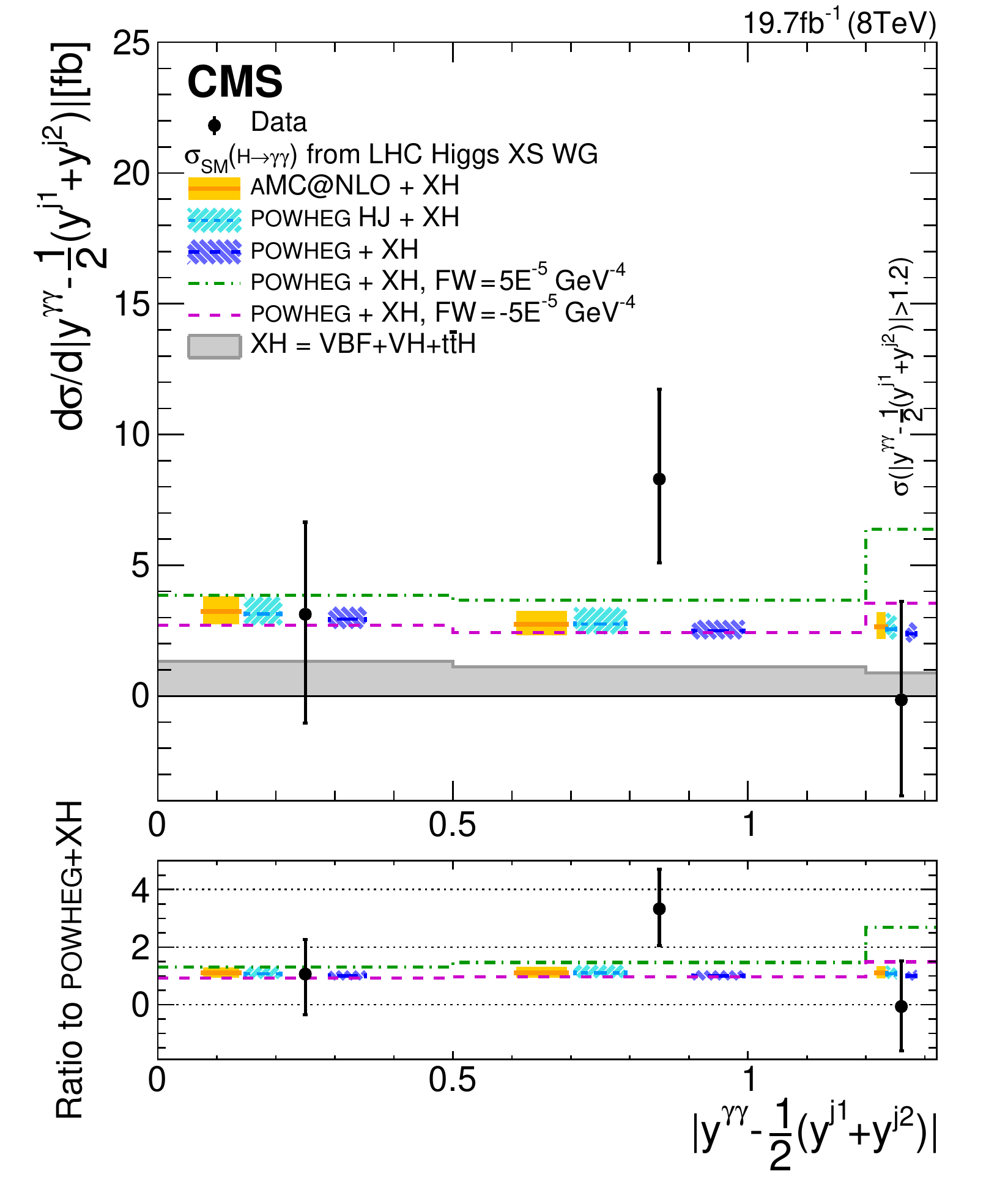}
   \includegraphics[width=0.45\textwidth,]{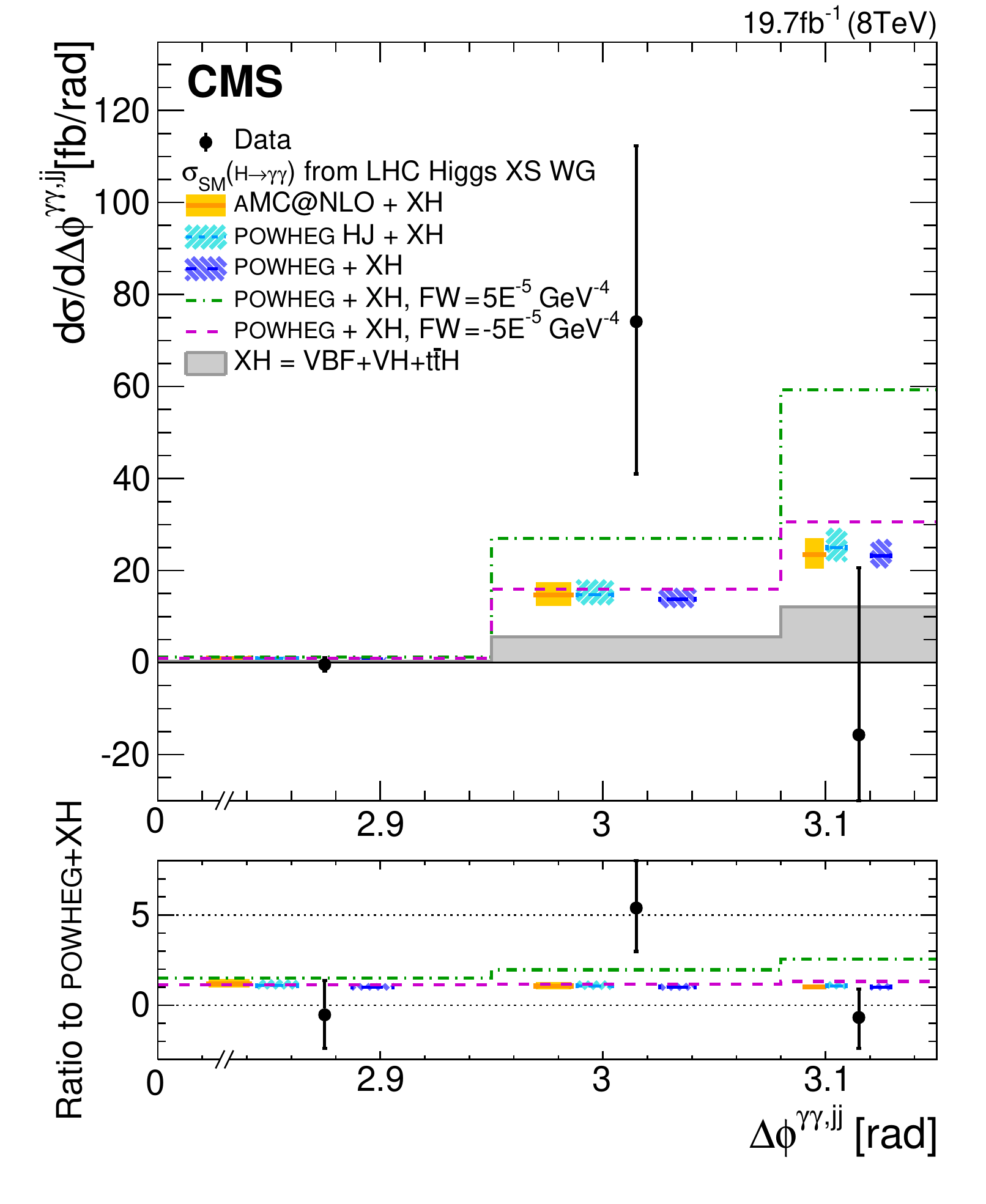}
   \caption{The $\Hgg$ differential cross section for H+2 jets events, with jets within $\abs{\eta}<4.7$, as a function of (upper left) $\deltaphijj$, (upper right) $\deltaetajj$, (lower left) Zeppenfeld variable, and (lower right) $\deltaphiggjj$. All the SM contributions are normalized to their cross section from Ref.~\cite{LHCHiggsCrossSectionWorkingGroup:2013}. Theoretical uncertainties in the renormalization and factorization scales, PDF, and branching fraction are added in quadrature. The error bars on data points reflect both statistical and systematic uncertainties. In the Zeppenfeld distribution, the last bin sums the events with values above 1.2. For each graph, the bottom panel shows the ratio of data to theoretical predictions from the $\POWHEG$ generator.}
  \label{fig:DataTheoryComp_2j_Eta47}
\end{figure*}

{\tolerance=1600
The measured differential cross sections observed in data, given for each bin by $\mu_i N^{\text{gen}}_{i}$, are compared with predictions for inclusive production in Fig.~\ref{fig:DataTheoryComp_Inclusive}, and for jet observables in Figs.~\ref{fig:DataTheoryComp_1j_Eta25} and \ref{fig:DataTheoryComp_2j_Eta47}.
The total theoretical uncertainty included in these comparisons is computed by adding in an uncorrelated way the uncertainties in the choice of PDF, renormalization and factorization scale, and the branching fraction. The uncertainties in the ggH mechanism, PDF choice, and the renormalization and factorization scales are computed with \HRES, \POWHEG, \POWHEGMINLO and \MADGRAPHAMCATNLO, as described above, while the branching fraction for all production mechanisms, as well as the scale and PDF for the associated production mechanisms are taken from Ref.~\cite{LHCHiggsCrossSectionWorkingGroup:2013}. Distributions for inclusive observables computed with \HRES, \POWHEG, and \MADGRAPHAMCATNLO, and jet-related observables computed with \POWHEG, \POWHEGMINLO and \MADGRAPHAMCATNLO, including latest higher-order corrections, are compatible within their uncertainties, and also compatible with the data.
\par}

Figure~\ref{fig:DataTheoryComp_Inclusive}~(upper left) shows the $\pt$ distribution of the Higgs boson, which is sensitive to higher-order corrections in perturbative QCD.
Figure~\ref{fig:DataTheoryComp_Inclusive}~(upper right) shows the absolute rapidity distribution of the Higgs boson, which is sensitive to the proton PDF, as well as to the production mechanism. Figure~\ref{fig:DataTheoryComp_Inclusive}~(lower left) shows the $\deltaphigg$ distribution.
Figure~\ref{fig:DataTheoryComp_Inclusive}~(lower right) displays the \costhetastar distribution, which is sensitive to the spin of the Higgs boson. The two spin-2 samples indicate deviations relative to the SM predictions. As in the case of Ref.~\cite{HggLegacy}, the data do not have sufficient sensitivity to discriminate between spin-2 and spin-0 hypotheses.

Figure~\ref{fig:DataTheoryComp_1j_Eta25}~(upper left) shows the $\pt$ distribution for the leading jet, which is sensitive to higher-order QCD effects, and Fig.~\ref{fig:DataTheoryComp_1j_Eta25}~(upper right) shows the rapidity difference between the Higgs boson and the leading jet. The distribution in the number of jets is displayed in Fig.~\ref{fig:DataTheoryComp_1j_Eta25}~(lower left). The last bin gives the cross section for signal events with at least three jets.
The distribution of the dijet mass shown in Fig.~\ref{fig:DataTheoryComp_1j_Eta25}~(lower right) is sensitive to the VBF production mechanism, and especially to a possible anomalous electroweak production in the tail of the distribution. Large contributions from new processes modifying triple gauge couplings would be detected as excesses either in $\ptgg$, $\ptja$ or in $\mjj$ distributions. The distributions in data are compatible with expectations from the SM within their uncertainties.
Figure~\ref{fig:DataTheoryComp_2j_Eta47}~(upper left) shows the difference in azimuthal angle $\deltaphijj$ between the two jets of highest $\pt$.
Figures~\ref{fig:DataTheoryComp_2j_Eta47}~(upper right),~\ref{fig:DataTheoryComp_2j_Eta47}~(lower left), and~\ref{fig:DataTheoryComp_2j_Eta47}~(lower right) show, respectively, the distribution of the rapidity difference between the two jets, the Zeppenfeld variable, and the azimuthal difference between the Higgs boson and the dijet system. These angular variables are sensitive to the VBF topology, and large contributions from anomalous couplings are not observed in data.
The distributions in data are compatible with SM predictions within their statistical, systematic, and theoretical uncertainties.

\section{Summary}
\label{sec:conclusions}

A measurement was carried out of differential cross sections as a function of kinematic observables in the \Hgg decay channel, using data collected by the CMS experiment at $\sqrt{s}=8\TeV$, corresponding to an integrated luminosity of 19.7\fbinv. The measurement was performed for events with two isolated photons in the kinematic range $\ptga/\mgg>1/3$ and $\ptgb/\mgg>1/4$, with photon pseudorapidities within $\abs{\eta}<2.5$.
Photon identification was chosen to reduce the dependence of the measurement on the kinematics of the signal. Event classification relied on an estimator of diphoton mass resolution. The signal extraction and the unfolding of experimental resolution were performed simultaneously in all bins of the chosen observables.
In this kinematic range, the fiducial cross section was measured to be $32 \pm 10$\unit{fb}.

The differential cross section of the Higgs boson was measured, inclusively in the number of jets, as a function of its transverse momentum $\ptgg$, its rapidity $\ygg$, the Collins--Soper angular variable \costhetastar, the difference in azimuthal angle between the two photons $\deltaphigg$, and the number of associated jets $\Njets$.
The transverse momentum of the leading jet $\ptja$, and the difference in rapidity between the Higgs boson and the leading jet $\yggj$ were determined in events with at least one accompanying jet.
In events with at least two jets, measurements were made of the dijet mass $\mjj$, the azimuth between the two jets $\deltaphijj$, the pseudorapidity difference between the two jets $\deltaetajj$, the Zeppenfeld variable $\Zepp$, and the azimuthal angle between the Higgs boson and the dijet system $\deltaphiggjj$.
The differential cross sections were compared with several SM and beyond SM calculations,
and found to be compatible with the SM predictions within statistical, systematic, and theoretical uncertainties.
With more data, anomalous couplings of the Higgs boson could be measured from its differential distributions. It would be possible to discriminate among different SM predictions for Higgs boson production those that are providing the best description.

\begin{acknowledgments}
\hyphenation{Bundes-ministerium Forschungs-gemeinschaft Forschungs-zentren} We congratulate our colleagues in the CERN accelerator departments for the excellent performance of the LHC and thank the technical and administrative staffs at CERN and at other CMS institutes for their contributions to the success of the CMS effort. In addition, we gratefully acknowledge the computing centers and personnel of the Worldwide LHC Computing Grid for delivering so effectively the computing infrastructure essential to our analyses. Finally, we acknowledge the enduring support for the construction and operation of the LHC and the CMS detector provided by the following funding agencies: the Austrian Federal Ministry of Science, Research and Economy and the Austrian Science Fund; the Belgian Fonds de la Recherche Scientifique, and Fonds voor Wetenschappelijk Onderzoek; the Brazilian Funding Agencies (CNPq, CAPES, FAPERJ, and FAPESP); the Bulgarian Ministry of Education and Science; CERN; the Chinese Academy of Sciences, Ministry of Science and Technology, and National Natural Science Foundation of China; the Colombian Funding Agency (COLCIENCIAS); the Croatian Ministry of Science, Education and Sport, and the Croatian Science Foundation; the Research Promotion Foundation, Cyprus; the Ministry of Education and Research, Estonian Research Council via IUT23-4 and IUT23-6 and European Regional Development Fund, Estonia; the Academy of Finland, Finnish Ministry of Education and Culture, and Helsinki Institute of Physics; the Institut National de Physique Nucl\'eaire et de Physique des Particules~/~CNRS, and Commissariat \`a l'\'Energie Atomique et aux \'Energies Alternatives~/~CEA, France; the Bundesministerium f\"ur Bildung und Forschung, Deutsche Forschungsgemeinschaft, and Helmholtz-Gemeinschaft Deutscher Forschungszentren, Germany; the General Secretariat for Research and Technology, Greece; the National Scientific Research Foundation, and National Innovation Office, Hungary; the Department of Atomic Energy and the Department of Science and Technology, India; the Institute for Studies in Theoretical Physics and Mathematics, Iran; the Science Foundation, Ireland; the Istituto Nazionale di Fisica Nucleare, Italy; the Ministry of Science, ICT and Future Planning, and National Research Foundation (NRF), Republic of Korea; the Lithuanian Academy of Sciences; the Ministry of Education, and University of Malaya (Malaysia); the Mexican Funding Agencies (CINVESTAV, CONACYT, SEP, and UASLP-FAI); the Ministry of Business, Innovation and Employment, New Zealand; the Pakistan Atomic Energy Commission; the Ministry of Science and Higher Education and the National Science Centre, Poland; the Funda\c{c}\~ao para a Ci\^encia e a Tecnologia, Portugal; JINR, Dubna; the Ministry of Education and Science of the Russian Federation, the Federal Agency of Atomic Energy of the Russian Federation, Russian Academy of Sciences, and the Russian Foundation for Basic Research; the Ministry of Education, Science and Technological Development of Serbia; the Secretar\'{\i}a de Estado de Investigaci\'on, Desarrollo e Innovaci\'on and Programa Consolider-Ingenio 2010, Spain; the Swiss Funding Agencies (ETH Board, ETH Zurich, PSI, SNF, UniZH, Canton Zurich, and SER); the Ministry of Science and Technology, Taipei; the Thailand Center of Excellence in Physics, the Institute for the Promotion of Teaching Science and Technology of Thailand, Special Task Force for Activating Research and the National Science and Technology Development Agency of Thailand; the Scientific and Technical Research Council of Turkey, and Turkish Atomic Energy Authority; the National Academy of Sciences of Ukraine, and State Fund for Fundamental Researches, Ukraine; the Science and Technology Facilities Council, UK; the US Department of Energy, and the US National Science Foundation.

Individuals have received support from the Marie-Curie program and the European Research Council and EPLANET (European Union); the Leventis Foundation; the A. P. Sloan Foundation; the Alexander von Humboldt Foundation; the Belgian Federal Science Policy Office; the Fonds pour la Formation \`a la Recherche dans l'Industrie et dans l'Agriculture (FRIA-Belgium); the Agentschap voor Innovatie door Wetenschap en Technologie (IWT-Belgium); the Ministry of Education, Youth and Sports (MEYS) of the Czech Republic; the Council of Science and Industrial Research, India; the HOMING PLUS program of the Foundation for Polish Science, cofinanced from European Union, Regional Development Fund; the OPUS program of the National Science Center (Poland); the Compagnia di San Paolo (Torino); the Consorzio per la Fisica (Trieste); MIUR project 20108T4XTM (Italy); the Thalis and Aristeia programs cofinanced by EU-ESF and the Greek NSRF; the National Priorities Research Program by Qatar National Research Fund; the Rachadapisek Sompot Fund for Postdoctoral Fellowship, Chulalongkorn University (Thailand); and the Welch Foundation, contract C-1845.
\end{acknowledgments}

\bibliography{auto_generated}

\appendix
\section{Appendix}

\begin{table*}[tp]
\topcaption{
Values of the $\Pp\Pp \to \Hgg$ differential cross sections as a function of kinematic observables as measured in data and as predicted in SM simulations.
For each observable the fit to \mgg is performed simultaneously in all the bins.
Since the signal mass is profiled for each observable, the best fit $\hat{m}_\PH$ varies from observable to observable.
}
\label{tab:xsec}
\centering
\resizebox*{!}{\dimexpr\textheight-3\baselineskip\relax}{
\begin{tabular}{l*{6}{c}}
\hline
 & {\centering $\sigma_{\text{obs}} [\text{fb}]$ { $(\hat{m}_\PH=124.4\GeV)$}}  & $\sigma_{\MADGRAPHAMCATNLO+\mathrm{XH}}$ & $\sigma_{\HRES+\mathrm{XH}}$ & $\sigma_{\POWHEG+\mathrm{XH}}$ &  \\
 \hline
 Fiducial cross section   & $32.2^{+10.1}_{-9.7}$  & $30.3^{+5.9}_{-4.7}$  & $31.5^{+4.0}_{-3.0}$  & $32.0^{+5.8}_{-4.6}$  &  \\[0.2ex]
\hline
\ptgg & {\centering $\sigma_{\text{obs}} [\text{fb}]$ {  $(\hat{m}_\PH=124.8\GeV)$}}  & $\sigma_{\MADGRAPHAMCATNLO+\mathrm{XH}}$ & $\sigma_{\HRES+\mathrm{XH}}$ & $\sigma_{\POWHEG+\mathrm{XH}}$ &  \\
\hline
  $0$--$15\GeV$ & $9.0^{+6.4}_{-6.2}$  & $7.5^{+1.5}_{-1.2}$  & $8.6^{+0.9}_{-0.9}$  & $8.6^{+1.6}_{-1.3}$  &  \\
  $15$--$26\GeV$ & $2.0^{+4.9}_{-5.5}$  & $5.9^{+1.2}_{-1.0}$  & $6.8^{+0.8}_{-0.7}$  & $6.6^{+1.3}_{-1.0}$  &  \\
  $26$--$43\GeV$ & $3.4^{+4.8}_{-4.6}$  & $6.1^{+1.3}_{-1.0}$  & $6.3^{+1.0}_{-0.7}$  & $6.2^{+1.1}_{-0.9}$  &  \\
 $43$--$72\GeV$ & $6.2^{+3.7}_{-3.5}$  & $5.2^{+1.1}_{-0.9}$  & $5.2^{+0.9}_{-0.6}$  & $5.1^{+0.9}_{-0.7}$  &  \\
 $72$--$125\GeV$ & $4.6^{+2.4}_{-2.7}$  & $3.4^{+0.7}_{-0.6}$  & $3.3^{+0.6}_{-0.4}$  & $3.3^{+0.6}_{-0.4}$  &  \\
  $125$--$200\GeV$ & $2.6^{+1.0}_{-1.0}$  & $1.4^{+0.3}_{-0.2}$  & $1.3^{+0.3}_{-0.3}$  & $1.3^{+0.2}_{-0.2}$  &  \\
 $>$200\GeV & $0.7^{+0.5}_{-0.4}$  & $0.6^{+0.1}_{-0.1}$  & $0.5^{+0.1}_{-0.1}$  & $0.6^{+0.2}_{-0.1}$  &  \\
\hline
$\abs{\costhetastar}$ & {\centering $\sigma_{\text{obs}} [\text{fb}]$ {  $(\hat{m}_\PH=124.7\GeV)$}}  & $\sigma_{\MADGRAPHAMCATNLO+\mathrm{XH}}$ & $\sigma_{\HRES+\mathrm{XH}}$ & $\sigma_{\POWHEG+\mathrm{XH}}$ &  \\
\hline
  $0.00$--$0.11$ & $4.9^{+3.3}_{-3.3}$  & $4.3^{+0.8}_{-0.7}$  & $4.8^{+0.5}_{-0.5}$  & $4.6^{+0.8}_{-0.7}$  &  \\
  $0.11$--$0.23$ & $3.3^{+3.2}_{-3.1}$  & $4.7^{+0.9}_{-0.7}$  & $5.2^{+0.7}_{-0.6}$  & $5.0^{+0.9}_{-0.7}$  &  \\
  $0.23$--$0.36$ & $11.7^{+3.8}_{-3.5}$  & $4.9^{+1.0}_{-0.8}$  & $5.4^{+0.7}_{-0.5}$  & $5.2^{+0.9}_{-0.7}$  &  \\
  $0.36$--$0.53$ & $2.4^{+5.6}_{-5.1}$  & $6.1^{+1.2}_{-1.0}$  & $6.7^{+0.9}_{-0.6}$  & $6.4^{+1.2}_{-0.9}$  &  \\
  $0.53$--$1.00$ & $7.3^{+7.3}_{-7.1}$  & $10.2^{+2.0}_{-1.6}$  & $9.6^{+1.4}_{-1.0}$  & $10.7^{+2.0}_{-1.5}$  &  \\
\hline
\deltaphigg & {\centering $\sigma_{\text{obs}} [\text{fb}]$ {  $(\hat{m}_\PH=124.6\GeV)$}}  & $\sigma_{\MADGRAPHAMCATNLO+\mathrm{XH}}$ & $\sigma_{\HRES+\mathrm{XH}}$ & $\sigma_{\POWHEG+\mathrm{XH}}$ &  \\
\hline
  $0.00$--$0.92$ & $0.9^{+0.9}_{-0.8}$  & $1.1^{+0.2}_{-0.2}$  & $0.9^{+0.1}_{-0.1}$  & $1.0^{+0.2}_{-0.1}$  &  \\
  $0.92$--$1.72$ & $2.6^{+1.4}_{-1.6}$  & $2.2^{+0.4}_{-0.3}$  & $2.0^{+0.3}_{-0.2}$  & $2.1^{+0.4}_{-0.3}$  &  \\
  $1.72$--$2.37$ & $2.2^{+2.6}_{-2.5}$  & $4.2^{+0.9}_{-0.7}$  & $4.0^{+0.7}_{-0.5}$  & $4.2^{+0.7}_{-0.5}$  &  \\
  $2.37$--$2.74$ & $8.2^{+4.1}_{-3.9}$  & $5.9^{+1.2}_{-1.0}$  & $5.9^{+0.9}_{-0.7}$  & $5.9^{+1.1}_{-0.8}$  &  \\
  $2.74$--$2.97$ & $2.8^{+4.4}_{-5.4}$  & $7.4^{+1.5}_{-1.2}$  & $8.2^{+1.1}_{-0.7}$  & $8.2^{+1.5}_{-1.2}$  &  \\
  $2.97$--$3.15$ & $14.4^{+9.0}_{-6.6}$  & $9.4^{+1.9}_{-1.5}$  & $10.5^{+1.0}_{-1.3}$  & $10.6^{+2.0}_{-1.6}$  &  \\
\hline
$\abs{\ygg}$  & {\centering $\sigma_{\text{obs}} [\text{fb}]$ {  $(\hat{m}_\PH=124.2\GeV)$}}  & $\sigma_{\MADGRAPHAMCATNLO+\mathrm{XH}}$ & $\sigma_{\HRES+\mathrm{XH}}$ & $\sigma_{\POWHEG+\mathrm{XH}}$ &  \\
\hline
  $0.00$--$0.16$ & $2.4^{+2.6}_{-2.6}$  & $3.1^{+0.6}_{-0.5}$  & $3.2^{+0.4}_{-0.3}$  & $3.2^{+0.6}_{-0.5}$  &  \\
  $0.16$--$0.35$ & $2.2^{+2.7}_{-2.6}$  & $3.7^{+0.7}_{-0.6}$  & $3.7^{+0.5}_{-0.4}$  & $3.9^{+0.7}_{-0.6}$  &  \\
  $0.35$--$0.58$ & $9.5^{+3.3}_{-3.2}$  & $4.3^{+0.8}_{-0.7}$  & $4.4^{+0.6}_{-0.5}$  & $4.5^{+0.8}_{-0.7}$  &  \\
  $0.58$--$0.90$ & $5.4^{+4.5}_{-3.7}$  & $5.6^{+1.1}_{-0.9}$  & $5.9^{+0.9}_{-0.6}$  & $5.8^{+1.1}_{-0.8}$  &  \\
  $0.90$--$2.50$ & $15.9^{+11.9}_{-11.1}$  & $13.8^{+2.7}_{-2.1}$  & $14.4^{+1.8}_{-1.3}$  & $14.7^{+2.7}_{-2.1}$  &  \\
\hline
\Njets  & {\centering $\sigma_{\text{obs}} [\text{fb}]$ {  $(\hat{m}_\PH=124.6\GeV)$}}  & $\sigma_{\MADGRAPHAMCATNLO+\mathrm{XH}}$ & $\sigma_{\POWHEG}$ $_{\mathrm{HJ+XH}}$ & $\sigma_{\POWHEG+\mathrm{XH}}$ &  \\
\hline
  $0$ & $20.2^{+9.3}_{-9.0}$  & $17.4^{+3.5}_{-2.8}$  & $18.1^{+6.0}_{-2.5}$  & $19.0^{+3.6}_{-2.8}$  &  \\
  $1$ & $4.3^{+6.4}_{-6.3}$  & $8.7^{+1.8}_{-1.5}$  & $9.6^{+2.2}_{-1.3}$  & $9.2^{+1.7}_{-1.3}$  &  \\
  $2$ & $3.9^{+3.7}_{-3.8}$  & $3.1^{+0.6}_{-0.5}$  & $3.1^{+0.7}_{-0.4}$  & $2.8^{+0.5}_{-0.3}$  &  \\
  $\geq$3 & $2.5^{+2.2}_{-2.2}$  & $1.1^{+0.2}_{-0.2}$  & $1.0^{+0.3}_{-0.1}$  & $1.0^{+0.2}_{-0.1}$  &  \\
\hline
\ptja &  {\centering $\sigma_{\text{obs}} [\text{fb}]$ {  $(\hat{m}_\PH=124.3\GeV)$}}  & $\sigma_{\MADGRAPHAMCATNLO+\mathrm{XH}}$ & $\sigma_{\POWHEG}$ $_{\mathrm{HJ+XH}}$ & $\sigma_{\POWHEG+\mathrm{XH}}$ &  \\
\hline
  $25$--$40\GeV$ & $2.1^{+6.8}_{-6.6}$  & $4.4^{+1.0}_{-0.8}$  & $4.5^{+1.2}_{-0.6}$  & $4.5^{+0.8}_{-0.6}$  &  \\
  $40$--$64\GeV$ & $6.4^{+5.6}_{-5.0}$  & $3.7^{+0.8}_{-0.7}$  & $3.9^{+0.9}_{-0.5}$  & $3.8^{+0.6}_{-0.5}$  &  \\
  $64$--$104\GeV$ & $3.4^{+3.4}_{-3.3}$  & $2.7^{+0.5}_{-0.4}$  & $2.9^{+0.6}_{-0.4}$  & $2.7^{+0.4}_{-0.3}$  &  \\
  $104$--$200\GeV$ & $0.8^{+1.9}_{-1.8}$  & $1.6^{+0.3}_{-0.2}$  & $1.8^{+0.4}_{-0.3}$  & $1.6^{+0.3}_{-0.2}$  &  \\
  $>$200\GeV & $0.8^{+0.8}_{-0.7}$  & $0.4^{+0.1}_{-0.1}$  & $0.5^{+0.1}_{-0.1}$  & $0.4^{+0.1}_{-0.1}$  &  \\
\hline
\yggj &  {\centering $\sigma_{\text{obs}} [\text{fb}]$ {  $(\hat{m}_\PH=124.5\GeV)$}}  & $\sigma_{\MADGRAPHAMCATNLO+\mathrm{XH}}$ & $\sigma_{\POWHEG}$ $_{\mathrm{HJ+XH}}$ & $\sigma_{\POWHEG+\mathrm{XH}}$ &  \\
\hline
  $0.00$--$0.62$ & $6.1^{+4.0}_{-3.8}$  & $3.8^{+0.7}_{-0.6}$  & $4.0^{+0.9}_{-0.5}$  & $3.7^{+0.6}_{-0.5}$  &  \\
  $0.62$--$1.25$ & $-1.1^{+4.4}_{-3.6}$  & $3.4^{+0.7}_{-0.6}$  & $3.6^{+0.8}_{-0.5}$  & $3.4^{+0.6}_{-0.5}$  &  \\
  $1.25$--$1.92$ & $5.2^{+3.0}_{-2.8}$  & $2.7^{+0.5}_{-0.5}$  & $2.9^{+0.7}_{-0.4}$  & $2.7^{+0.5}_{-0.4}$  &  \\
  $>$1.92 & $1.6^{+4.1}_{-4.8}$  & $3.0^{+0.6}_{-0.5}$  & $3.2^{+0.7}_{-0.5}$  & $3.1^{+0.5}_{-0.4}$  &  \\
\hline
\mjj &  {\centering $\sigma_{\text{obs}} [\text{fb}]$ {  $(\hat{m}_\PH=124.7\GeV)$}}  & $\sigma_{\MADGRAPHAMCATNLO+\mathrm{XH}}$ & $\sigma_{\POWHEG}$ $_{\mathrm{HJ+XH}}$ & $\sigma_{\POWHEG+\mathrm{XH}}$ &  \\
\hline
  $0$--$153\GeV$ & $6.8^{+4.5}_{-4.0}$  & $3.0^{+0.7}_{-0.6}$  & $2.9^{+0.7}_{-0.4}$  & $2.6^{+0.4}_{-0.3}$  &  \\
  $153$--$455\GeV$ & $1.8^{+3.0}_{-3.2}$  & $2.1^{+0.4}_{-0.4}$  & $2.1^{+0.5}_{-0.3}$  & $2.0^{+0.3}_{-0.2}$  &  \\
  $455$--$1000\GeV$ & $0.3^{+1.2}_{-1.0}$  & $0.8^{+0.1}_{-0.1}$  & $0.9^{+0.1}_{-0.1}$  & $0.8^{+0.1}_{-0.1}$  &  \\
  $>$1000\GeV & $-0.2^{+0.3}_{-0.2}$  & $0.3^{+0.0}_{-0.0}$  & $0.3^{+0.0}_{-0.0}$  & $0.3^{+0.0}_{-0.0}$  &  \\
\hline
\deltaetajj & {\centering $\sigma_{\text{obs}} [\text{fb}]$ {  $(\hat{m}_\PH=124.4\GeV)$}}  & $\sigma_{\MADGRAPHAMCATNLO+\mathrm{XH}}$ & $\sigma_{\POWHEG}$ $_{\mathrm{HJ+XH}}$ & $\sigma_{\POWHEG+\mathrm{XH}}$ &  \\
\hline
  $0.00$--$1.65$ & $3.0^{+3.6}_{-3.5}$  & $2.9^{+0.6}_{-0.5}$  & $2.8^{+0.7}_{-0.4}$  & $2.5^{+0.4}_{-0.3}$  &  \\
  $1.65$--$4.30$ & $3.7^{+2.8}_{-2.6}$  & $2.4^{+0.5}_{-0.4}$  & $2.3^{+0.5}_{-0.3}$  & $2.2^{+0.3}_{-0.3}$  &  \\
  $4.30$--$10.00$ & $0.1^{+1.0}_{-0.9}$  & $0.9^{+0.1}_{-0.1}$  & $0.9^{+0.1}_{-0.1}$  & $0.9^{+0.1}_{-0.1}$  &  \\
\hline
\deltaphiggjj  & {\centering $\sigma_{\text{obs}} [\text{fb}]$ {  $(\hat{m}_\PH=124.5\GeV)$}}  & $\sigma_{\MADGRAPHAMCATNLO+\mathrm{XH}}$ & $\sigma_{\POWHEG}$ $_{\mathrm{HJ+XH}}$ & $\sigma_{\POWHEG+\mathrm{XH}}$ &  \\
\hline
  $0.00$--$2.95$ & $-0.1^{+0.2}_{-0.2}$  & $0.1^{+0.0}_{-0.0}$  & $0.1^{+0.0}_{-0.0}$  & $0.1^{+0.0}_{-0.0}$  &  \\
  $2.95$--$3.08$ & $9.6^{+5.0}_{-4.3}$  & $1.9^{+0.4}_{-0.3}$  & $1.9^{+0.4}_{-0.3}$  & $1.8^{+0.3}_{-0.2}$  &  \\
  $3.08$--$3.15$ & $-1.1^{+2.5}_{-2.8}$  & $1.6^{+0.2}_{-0.2}$  & $1.7^{+0.3}_{-0.2}$  & $1.6^{+0.2}_{-0.2}$  &  \\
\hline
\deltaphijj & {\centering $\sigma_{\text{obs}} [\text{fb}]$ {  $(\hat{m}_\PH=124.6\GeV)$}}  & $\sigma_{\MADGRAPHAMCATNLO+\mathrm{XH}}$ & $\sigma_{\POWHEG}$ $_{\mathrm{HJ+XH}}$ & $\sigma_{\POWHEG+\mathrm{XH}}$ &  \\
\hline
  $0.00$--$0.88$ & $1.5^{+2.0}_{-1.9}$  & $1.6^{+0.3}_{-0.3}$  & $1.7^{+0.3}_{-0.2}$  & $1.4^{+0.2}_{-0.2}$  &  \\
  $0.88$--$1.82$ & $1.1^{+2.1}_{-2.3}$  & $1.7^{+0.3}_{-0.3}$  & $1.6^{+0.3}_{-0.2}$  & $1.6^{+0.3}_{-0.2}$  &  \\
  $1.82$--$3.15$ & $5.0^{+3.8}_{-3.5}$  & $3.0^{+0.6}_{-0.5}$  & $2.8^{+0.7}_{-0.4}$  & $2.7^{+0.4}_{-0.3}$  &  \\
\hline
\Zepp  & {\centering $\sigma_{\text{obs}} [\text{fb}]$ {  $(\hat{m}_\PH=124.6\GeV)$}}  & $\sigma_{\MADGRAPHAMCATNLO+\mathrm{XH}}$ & $\sigma_{\POWHEG}$ $_{\mathrm{HJ+XH}}$ & $\sigma_{\POWHEG+\mathrm{XH}}$ &  \\
\hline
  $0.0$--$0.5$ & $1.6^{+1.8}_{-2.1}$  & $1.6^{+0.3}_{-0.2}$  & $1.6^{+0.3}_{-0.2}$  & $1.5^{+0.2}_{-0.2}$  &  \\
  $0.5$--$1.2$ & $5.8^{+2.4}_{-2.2}$  & $1.9^{+0.4}_{-0.3}$  & $1.9^{+0.4}_{-0.2}$  & $1.7^{+0.3}_{-0.2}$  &  \\
  $>$1.2 & $-0.1^{+3.8}_{-3.7}$  & $2.6^{+0.6}_{-0.5}$  & $2.6^{+0.6}_{-0.3}$  & $2.4^{+0.4}_{-0.3}$  &  \\
\hline
\end{tabular}
}
\end{table*}
\cleardoublepage \section{The CMS Collaboration \label{app:collab}}\begin{sloppypar}\hyphenpenalty=5000\widowpenalty=500\clubpenalty=5000\textbf{Yerevan Physics Institute,  Yerevan,  Armenia}\\*[0pt]
V.~Khachatryan, A.M.~Sirunyan, A.~Tumasyan
\vskip\cmsinstskip
\textbf{Institut f\"{u}r Hochenergiephysik der OeAW,  Wien,  Austria}\\*[0pt]
W.~Adam, E.~Asilar, T.~Bergauer, J.~Brandstetter, E.~Brondolin, M.~Dragicevic, J.~Er\"{o}, M.~Flechl, M.~Friedl, R.~Fr\"{u}hwirth\cmsAuthorMark{1}, V.M.~Ghete, C.~Hartl, N.~H\"{o}rmann, J.~Hrubec, M.~Jeitler\cmsAuthorMark{1}, V.~Kn\"{u}nz, A.~K\"{o}nig, M.~Krammer\cmsAuthorMark{1}, I.~Kr\"{a}tschmer, D.~Liko, T.~Matsushita, I.~Mikulec, D.~Rabady\cmsAuthorMark{2}, B.~Rahbaran, H.~Rohringer, J.~Schieck\cmsAuthorMark{1}, R.~Sch\"{o}fbeck, J.~Strauss, W.~Treberer-Treberspurg, W.~Waltenberger, C.-E.~Wulz\cmsAuthorMark{1}
% ...
\vskip\cmsinstskip
% ...
1:~~Also at Vienna University of Technology, Vienna, Austria\\
2:~~Also at CERN, European Organization for Nuclear Research, Geneva, Switzerland\\
% ...

\end{sloppypar}
\end{document}